\documentclass[12pt]{article}
\usepackage{amsfonts}

\input{tcilatex}

\begin{document}

\title{\textbf{The Pivotal Role of Causality in Local Quantum Physics}~}
\author{Bert Schroer \\
Institut f\"{u}r Theoretische Physik\\
FU-Berlin, Arnimallee 14, 14195 Berlin, Germany\\
presently: CBPF, Rua Dr. Xavier Sigaud, 22290-180 Rio de Janeiro, Brazil \\
schroer@cbpfsu1.cat.cbpf.br}
\date{March 1999\\
Dedicated to Professor Eyvind H. Wichmann on the occasion of his seventieth
birthday.\\
submitted to Journal of Physics A}
\maketitle

\begin{abstract}
In this article an attempt is made to present very recent conceptual and
computational developments in QFT as new manifestation of of old well
establihed physical principles. The vehicle for converting the
quantum-algebraic aspects of local quantum physics into more classical
geometric structures is the modular theory of Tomita. As the above named
laureate together with his collaborator showed for the first time in
sufficient generality, its use in physics goes through Einstein causality.
This line of research recently gained momentum when it was realized that it
is not only of great structural and conceptual innovative power (see section
4), but also promises a new computational road into nonperturbative QFT
(section 5) which, picturesquely speaking, enters the subject on the extreme
opposite (noncommutative) side relative to (Lagrangian) quantization.
\end{abstract}

\tableofcontents

\section{\protect\large Introduction}

Among the fundamental physical principles of this century which have stood
their ground in the transition from classical into quantum physics,
relativistic causality as well as the closely related locality of quantum
operators (together with the localization of quantum states) is certainly
the most prominent one.

This principle entered physics through Einsteins 1905 special relativity,
which in turn resulted from bringing the Galilei relativity principle of
classical mechanics into tune with Maxwell's theory of electromagnetism.
Therefore it incorporated Faraday's ``action at a neighborhood'' principle
which revolutionized 19$^{th}$ century physics.

The two different aspects of Einstein's special relativity, namely
Poincar\'{e} covariance and the locally causal propagation of waves in
Minkowski space were kept together in the classical setting. In the
adaptation of relativity to LQP (local quantum physics\footnote{%
We use this terminology, whenever we want to make clear that we relate the
principles of QFT with a different formalism than that based on quantization
through Lagrangian formalism.}) on the other hand \cite{Haag}, it is
appropriate to keep them at least initially apart in the form of positive
energy representations of the Poincar\'{e} group (leading to Wigner's
concept of particles) and Einstein causality of local observables (leading
to observable local fields and local generalized ``charges''). Here a
synthesis is also possible, but it happens on a deeper level than in the
classical setting and results in LQP as a new physical realm which is
conceptually very different from both classical field theory and general QT
(quantum theory). The elaboration of this last point constitutes one of the
aims of these notes. We will pay particular attention to those aspects of
LQP which are not within the reach of standard quantum physical intuition.

The most remarkable aspect of QFT in its more than 60 years existence in
addition to its great descriptive and computational success in perturbative
QED and the standard model, is certainly the perseverance of its causality
principle. In addition to the experimental support through the validity of
the Kramers-Kronig dispersion relations in high energy collisions up to the
shortest accessible distances, it is also the various unsuccessful
theoretical attempts to construct viable nonlocal theories\footnote{%
The meaning of ``nonlocal'' in these notes is not that of extended charged
objects in a theory of local observables (example: semiinfinite string like
spatial extensions of anyons or plektons in d=1+2 in order to support their
abelian/nonabelian braid group statistics), but rather refers to
hypothetical theories which have a fundamental cut-off or elementary length
in their algebra of observables.} which testify to the strength of this
principle. Despite intense efforts and much talk, nobody has succeeded to
construct a viable \textit{nonlocal} theory. The cutoff in Feynman-like
integrals or in euclidean functional integrals (which violate the
prerequisites for continuability to real time LQP) introduced by
phenomenologists in order to combat the apparent ``bad'' short distance
behavior stemming from perturbative causality down to arbitrary small
distances (which threaten the mathematical existence of models) are no
substitute for a conceptual analysis whether a viable nonlocal theory with
an elementary length which maintains a particle interpretation is possible
at all\footnote{%
A good antidote against speculations or light-hearted attitudes that e.g.
rotational invariant euclidean cutoffs (or any other kind of cutoff which
formally can be expected to maintain Lorentz covariance) could define a
consistent nonlocal real time theory, is to try to introduce one into one of
the exactly solvable d=1+1 factorizing models.}. Here ``viable'' is more
than mere mathematical existence, it is meant in the physical sense of
conceptual completeness. One requires that a theory is required to \textit{%
contain its own physical interpretation} i.e. that one does not have to
invent or borrow formulas from outside this theory as it is done in e. g.
phenomenological ``effective'' QFT. In the latter case most formulas linking
the calculations with measurable quantities cannot be derived or justified,
but as exprected in a phenomenological approach, have to be taken from a
more complete and fundamental framework. In addition ``effective ''
indicates that objects with this prefix as Lagrangians, actions etc. should
be dealt with different rules as those in renormalized perturbation theory.
On the other hand in a complete framework as LQP, one cannot only derive
(LSZ) scattering formulas which constitute an important aspect of particle
interpretation, but one can also obtain the composition laws of charges,
analytic and crossing properties of fields in particle states etc; in fact
there is presently no important structural or epistomological property which
the principles of LQP cannot address or account for. Only if it comes to
quantitative understanding of particle interaction processes one has to
resort to specific models, even though their full control is often very
problematic as a result of absense of systematic and reliable
nonperturbative methods.

Contrary to statements one sometimes finds in the literature, there is no
nonlocal Poincar\'{e} covariant scheme known, which guaranties the existence
of a time dependent (or its stationary reformulation) scattering formalism
together with the analytic and crossing properties of matrix-elements of the
S-operator and formfactors of local fields, and therefore could be used in
particle physics. Hence the importance of causality is also highlighted by
the failure of nonlocal modifications and the conspicuous absense of
physically viable alternatives. It is quite instructive to briefly look at
some of the more prominent failed attempts.

Already in the 50$^{ies}$ there were proposals to inject nonlocal aspects
through extended interaction-vertices in Lorentz invariant Lagrangians. As
mentioned before, this was motivated by the hope that a milder perturbative
short distance behavior in correlation functions may be helpful for
demonstrating the mathematical existence of the theory. It was soon
realized, that if one pursues the effect of such modifications up to
infinite order in perturbation theory, these nonlocal vertices would wreck
even macrocausality so that the theory looses its physical interpretation
alltogether. A similar fate occurred to the later proposal of Lee and Wick 
\cite{Lee-Wick} to allow for complex (+ complex conjugate, in order to
maintain hermiticity) poles in Feynman rules; it led to unacceptable time
precursors \cite{Ma-Swie}. In the last section we will present some results
on a new nonperturbative framework which incorporates and explains all the
results obtained insofar on explicit non-Lagrangian low-dimensional model
constructions. The very concepts of this approach use causality and locality
in a much more essential way than the various quantization approaches and in
addition this method throws considerable doubt in the belief that the
perturbative link between good short distance behavior and existence of the
theory has general validity.

Often the renormalization group ideas are used to justify a physical cutoff
with the hope that by softening short-distance behavior the model becomes
mathematically better defined and managable. But physical principles should
receive their limitation, as it always happened in the past, from other more
general principles and not from parameters into which one tries to dump ones
lack of knowledge about the mathematical existence of the theory within the
presently known principles. A phenomenological successful parameter with
fixed computational prescriptions is by itself is not a substitute for a
physical principle. Physical reality may unfold itself like an onion with
infinitely many layers of ever more general physical principles towards the
small, but it should still be possible to have a mathematically consistent
theory in each layer which is faithful to the principles valid in that
layer. This has been fully achieved for quantum mechanics, but this goal was
not yet reached in QFT as a result of lack of nontrivial d=1+3 models or
structural arguments which could demonstrate that the requirements allow for
nontrivial solutions. Even the recently emphasized duality between
asymptotically small/large coupling parameters only resulted in the
rephrasing of the problem to: \textit{does there exist a QFT which pocesses
these two asymptotes.} The existence problem of interacting QFT's in d=1+3
which persists to present times sets QFT apart from any other physical
theory as QM, Stat. Mech. or classical particles/field theories. In all
those cases one has explicite examples as well as proofs that the ``axioms''
are consistent with nontrivial dynamics. In this context one should note
that lattice theories define a different (mathematically easier) framework
which, if suitably restricted, shares with QFT that it is conceptually
complete as far as the notion of particle excitations and their scattering
theory (based on cluster properties) is concerned. In fact the correlation
functions of lattice algebras are expected to converge towards those of a
QFT in an appropriately defined scaling limit. Despite some control of the
extremely difficult scaling limits in certain special models as the d=2
Ising like models, the relation between the two theories remains largely not
understood.

Recently there was a more sophisticated attempt to go beyond the causal
setting of LQP via the use of noncommutative space time \cite{DFR}, based on
spatial uncertainty relations following from a quasiclassical quantization
interpretation of Einstein's field equation of general relativity and the
assumed absence of very small black holes (similar uncertainty relation for
the complete set of coordinates and momenta (i.e. for phase space) have been
postulated on the basis of string theory \cite{Veneziano}). These proposals,
especially if they are backed up by uncertainty relations whose derivation
is carried out in the spirit of Bohr-Rosenfeld as in \cite{DFR}, are not
that easily dismissed as the two previous ones. Such attempts do not just
try to graft cutoffs or elementary length onto the standard (Lagrangian,
functional integral) local framework, but rather are receptive to more
radical changes of the fundamentals of QFT. It is not easy to confront such
speculative new ideas with LQP, because it is more difficult to physically
interpret in such unusual frameworks than it is to rule out implanting
cutoffs into the standard framework. Whereas it is easy to agree that
sufficent intelligent noncommutative spacetime proposals may serve as
interesting tests for exploring the unknown territory beyond the reign of
Einstein causality, they are still far from being models for the elusive
``Quantum Gravity'', since they only replace the classical spacetime
indexing of nets with a noncommutative one. However any step beyond the
present causal framework must reobtain Einstein causality as an limiting
statement within some yet unknown new physical principle. Recently there
have been a lot of promises on the basis of string theory. But unfortunately
string theory, even aside from the total lack of experimental motivation,
had hardly added anything to conceptual problems despite its undeniable
mathematical enrichments. In fact in its present state it is mainly a loose
set of calculational recipes which suffer from a very unfortunate preference
of formalism over conceptual clarifications. Whereas LQP allows an intrinsic
characterization (e.g. in terms of correlation functions or observable nets)
independent on the way they have been manufactured (e.g. Lagrangian
quantization, bootstrap-formfactor method in d=1+1), string theory in its
more than 20 years of existence has not led to objects with an intrinsic
meaning independent of the computational rules (in addition to its
experimental invulnerability after it changed interpretation from the old
string theory of the dual model for strong interaction at laboratory
energies to an alleged theory of quantum gravitation thus jumping 15 orders
of magnitude. On the theoretical side such fundamental questions whether
strings are localized objects in spacetime (as the name seems to indicate)
or if the name is a short hand notation for specific spectral features have
nor yet been settled. Whereas admittedly many of the the popular
formulations of QFT based on canonical or functional integral quantization
start also from extrinsic formal requirements which in most cases cannot be
maintained after renormalization\footnote{%
Apart from some less interesting superrenormalizable models, the physically
meaningful renormalizable objects (which are also the only ones with a
chance of mathematical existence) are neither canonical nor representable by
functional integrals, but still fulfill the property of Einstein causality
together with certain spectral properties. The so-called ``causal
perturbation theory'' (see later) furnishes a more harmonious intrinsic
formulation for which the initial requirements are also reflected in the
results, and not only as a ``catalyzer'' of the mind.}, there exists at
least various intrinsic formulations.

Causality and locality are in a profound way related to the foundations of
quantum theory in the spirit of von Neumann. In von Neumann's formulation,
observables are represented by selfadjoint operators and measurements are
compatible if the operators commute. The totality of all measurements which
are relatively compatible with a given set (i.e. noncommutativity within
each set is allowed) generate a subalgebra: the commutant $\mathcal{L}%
^{\prime }$ of the given set of operators $\mathcal{L}$. In LQP, a
conceptual framework which was not yet available to von Neumann, one is
dealing with an isotonic ``net'' of subalgebras (in most physically
interesting cases von Neumann factors, i.e.with trivial center) $\mathcal{O}%
\rightarrow \mathcal{A}(\mathcal{O}),$ such that unlike quantum mechanics,
the spatial localization and the time duration of observables becomes an
integral part of the formalism. \textit{Causality gives an a-priori
information about the size} \textit{of spacetime }$\mathcal{O}$\textit{\
-affiliated operator von Neumann algebras:} 
\begin{equation}
\mathcal{A}(\mathcal{O})^{\prime }\supset \mathcal{A}(\mathcal{O}^{\prime })
\label{Einstein}
\end{equation}
in words: the commutant of the totality of local observables localized in
the spacetime region $\mathcal{O}$ contains the observables localized in its
spacelike complement (disjoint) $\mathcal{O}^{\prime }.$ In fact in most of
the cases the equality sign will hold in which case one calls this
strengthened (maximal) form of causality ``Haag duality'' \cite{Haag}\cite
{Sch Lec}: 
\begin{equation}
\mathcal{A}(\mathcal{O})^{\prime }=\mathcal{A}(\mathcal{O}^{\prime })
\end{equation}
In words, the spacelike localized measurements are not only commensurable
with the given observables in $\mathcal{O}$, but every measurement which is
commensurable with all observables in $\mathcal{O},$ is necessarily
localized in the causal complement $\mathcal{O}^{\prime }.$ Here we extended
for algebraic convenience von Neumann's notion of observables to the whole
complex von Neumann algebra generated by hermitian operators localized in $%
\mathcal{O}.$ If one starts the theory from a net indexed by compact regions 
$\mathcal{O}$ as double cones, then algebras associated with unbounded
regions $\mathcal{O}^{\prime }$ are defined as the von Neumann algebra
generated by all $\mathcal{A}(\mathcal{O}_{1})$ if $\mathcal{O}_{1}$ ranges
over all net indices $\mathcal{O}_{1}\subset \mathcal{O}^{\prime }.$

Whereas the Einstein causality (\ref{Einstein}) allows a traditional
formulation in terms of pointlike fields $A(x)$ as 
\begin{equation}
\left[ A(x),A(y)\right] =0,\,\,\,\left( x-y\right) ^{2}<0,
\end{equation}
Haag duality can only be formulated in the algebraic net setting of LQP.
This aspect is shared by many important properties and results presented in
this article. LQP is much more than a Teutonic pastime of reformulating
properties of fields in terms of algebraic properties of nets as one
realizes immediately if one looks into Haag's book.

One can prove that that Haag duality always holds after a suitable extension
of the net to the so-called dual net $\mathcal{A}(\mathcal{O})^{d}.$ The
latter may be defined independent of locality in terms of relative
commutation properties as 
\begin{equation}
\mathcal{A}(\mathcal{O})^{d}:=\bigcap_{\mathcal{O}_{1},\mathcal{O}%
_{1}^{\prime }\subset \mathcal{O}}\mathcal{A}(\mathcal{O}_{1})^{\prime }
\end{equation}
It is easy to check that the dual net is relatively local to the original
net 
\begin{equation}
A(\mathcal{O}_{1})\subset (\mathcal{A}(\mathcal{O})^{d})^{\prime },\,\,\,%
\mathcal{O}_{1}\subset \mathcal{O}^{\prime }
\end{equation}
in fact it is the maximal net relatively local to $\mathcal{A}(\mathcal{O}).$
Repeating this process, one obtains $\mathcal{A}(\mathcal{O})^{d}\subset 
\mathcal{A}(\mathcal{O})^{dd}$ and $\mathcal{A}(\mathcal{O})^{d}=\mathcal{A}(%
\mathcal{O})^{ddd}.$ Causality of the original net then means $\mathcal{A}(%
\mathcal{O})\subset \mathcal{A}(\mathcal{O})^{d},$ and therefore also $%
\mathcal{A}(\mathcal{O})^{dd}\subset \mathcal{A}(\mathcal{O})^{d}=\mathcal{A}%
(\mathcal{O})^{ddd}.$ It is costumary to use the word locality instead of
causality if one allows field algebras which involve fermionic structures.
Local algebras retain all of the mathematical properties of observable
algebras in that they contain no local annihilators. \ The extension by
charged objects with braid group statistics (only possible in spacetime
dimension d\TEXTsymbol{<}1+3) may lead to algebras (acting in a larger
Hilbert space) with weaker locality properties and the appearance of local
annihilators. Such objects are called ``localizable'' since they maintain
their \textit{relative locality} with respect to the neutral observable
subalgebra. The causal disjoint of the region of relative commutation is the
localization region of these charged operators. These considerations show
that causality, locality and localization in LQP have a close relation to
the notion of compatibility of measurements. The fundamental reason for all
such modifications in the interpretation of LQP versus QM is the different
structure of local algebras: the vacuum is not a pure state with respect to
any algebra which is contained in an $\mathcal{A}(\mathcal{O})$ with $%
\mathcal{O}^{\prime }$ nonempty, and the sharply localized algebras $%
\mathcal{A}(\mathcal{O})$ do not admit any pure states at all! Since these
fine points can only be appreciated with some more preparation, I will
postpone their presentation. Note that the quantization approach to QFT
based on the use of classical actions in euclidean functional integrals (and
the subsequent use of analytic continuation to get back to real spacetime)
is a global attempt to characterize vacuum expectation values of a would be
theory. The classical locality in the sense of local polynomial expressions
in fields and derivatives has no direct conceptual relation with the real
time locality in the above sense; in fact the analytically continued
``fields'' in the euclidean points are extremely nonlocal relatively with
respect to the real time fields. Unlike in statistical mechanics It does not
make sense to restrict the euclidean integration to localized configurations
with local supports since this has nothing to do with the localization of
real time physics where one may restrict states to localized subalgebras.
Nevertheless there are sufficient conditions under which the euclidean
correlation functions do permit to define models of real time QFT.

If the vacuum net is Haag dual, then all associated ``charged'' nets share
this property, unless the charges are nonabelian; in which case the
deviation from Haag duality is measured by the Jones index of the above
inclusion, or in physical terms the statistical- or quantum-dimension. If
even the vacuum representation violates Haag duality, this indicates
spontaneous symmetry breaking \cite{Roberts} i.e. not all internal symmetry
algebraic automorphisms are spatially implementable. As already mentioned,
in that case one can always maximize the local algebra to the dual algebras $%
\mathcal{A}^{d}(\mathcal{O}$) without destroying causality and without
changing the Hilbert space and in this way Haag duality is restored
(``essential duality''). This turns out to be related to the descend to the
unbroken part of the symmetry which allows (since it is a subgroup) more
invariants i.e. more observables. Although these matters are good
illustrations of the pivotal role of causality, we will concentrate on the
closely related modular properties of causal nets which will make their
appearance in the next section. QM does not know these concepts at all,
trying to add them would mean leaving QM, since their realization requires
infinite degrees of freedom.

Another structurally significant deviation is expected to result from the
fact that the vacuum becomes a thermal state with respect to the local
algebras $\mathcal{A}(\mathcal{O}).$ There are two different mechanisms
which generate thermal states: the coupling with a heat bath and the
thermality through restriction or localization and the creation of horizons.
The latter is in one class with with the Hawking-Unruh mechanism; the
difference being that in the localization situation the horizon is not
classical i.e. is not defined in terms of a differential geometric Killing
generator of a symmetry transformation of the metric.

Since the algebras of the type $\mathcal{A}(\mathcal{O})$ do not possess
pure states, the $\mathcal{O}/\mathcal{O}^{\prime }$ situation is totally
different from the tensor product factorization in terms of the quantization
box inside/outside in QM. In order to get back to a tensor product situation
and be able to apply the concepts of entanglement and entropy, one has to do
a sophisticated split which is only possible if one allows for a ``collar''
(see later) between $\mathcal{O}$ and $\mathcal{O}^{\prime }$. These
considerations show that certain things which one takes for granted as
properties of general QT actually loose their validity in LQP.

Since the thermal aspects of localization are analogous to those of black
holes, there is no chance to directly measure such tiny effects. However in
conceptual problems, e.g. the question if and how not only classical
relativistic field theory but also QFT excludes superluminal velocities,
these subtle differences play a crucial role. Imposing the usual algebraic
structure of QM onto the theory of photons will lead to nonsensical results.
Most sensational theoretical observations on causality violations which are
not allready wrong on a classical level suffer from incorrect tacit
assumptions.We urge the reader to read the reference \cite{Yng-Buch} and
also look at the source for that rebuttal.

Historically the first conceptually clear definition of localization of
relativistic wave function was given by Newton and Wigner \cite{New Wig} who
adapted Born's x-space probability interpretation to the Wigner relativistic
particle theory. Apparently the result that there is no exact satisfactory
relativistic localization (but only one sufficient for all practical
purposes), disappointed Wigner so much that he became distrustful of the
consistency of QFT in particle physics alltogether (private communication by
R. Haag). Whereas we know that this distrust was unjustified, we should at
the same time acknowledge Wigner's stubborn insistence in the importance of
the locality concept as a indispensable particle physics requirement in
addition the positive energy property and irreducibility of his
representations theory. Modular localization of subspaces of the Hilbert
space and of subalgebras on the other hand are not related to the Born
probability interpretation. Rather modular localized state vectors preempt
the existence of causally localized observables and have no counterpart at
all in N-particle quantum mechanics. As will be explained later modular
localization may serve as a starting point for the construction of
interacting nonperturbative LQP's \cite{Sch Lec}\cite{Sch AOP}\footnote{%
In fact the good modular localization properties of positive energy
properties, with the exception of Wigner's infinite component ``continuous
spin'' representations, are guarantied. Only in the infinite component case
it is not possible to come from the wedge localization to the spacelike cone
localization which is the coarsest localization from which one can still
obtain a Wigner particle interpretation.}. It is worthwhile to emphasize
that sharper localization of local algebras in LQP is not defined in terms
of smaller support properties of classical smearing functions of smeared
fields but rather in terms of intersection of algebras; although in many
cases as CCR- or CAR-algebras (or more generally Wightman fields) the
algebraic formulation (\ref{Einstein}) can be reduced to this more classical
concept.

Since the modular structure is in a deep way related to thermal behavior, it
is not surprising that the issue of thermality is also related with
localization. In fact as mentioned before, there are two manifestations of
thermality, the standard heat bath thermal behavior which is described by
Gibbs formula (or after having performed the thermodynamic limit by the KMS
condition), and thermality caused by localization either with classical
bifurcated Killing-horizons as in black holes \cite{Sew}, or in a purely
quantum manner as the boundary of the Minkowski space wedges or double
cones. In the latter case the KMS state has no natural limiting description
in terms of a Gibbs formula (which only applies to type I and II, but not to
type III von Neumann algebras), a fact which is also related to the fact
that the hamiltonian (of the ground state problem) is bounded from below,
whereas the e.g. Lorentz boost (the modular operator of the wedge algebra in
the vacuum state) is not \cite{Sch AOP}. In \cite{Jaekel} the reader also
finds an discussion of localization and cluster properties in a heat bath
thermal state. In these notes we will not enter these interesting thermal
aspects. Recent results indicate that the division between heat bath- and
localization-thermality may not be as sharp as it appears at first sight 
\cite{Sch-Wie2}

\section{\protect\large Locality and Free Particles}

The best way to make the pivotal nature of causality manifest, is to enter
QFT via Wigner's group theoretical characterization of particles by
irreducible positive energy representations with good localization
properties. It is well known that the Wigner wave functions $\psi $ of
massive spin s particles have 2s+1 components and (differently from
covariant fields) transform in a manifestly unitary but p-dependent way: 
\begin{equation}
(U(\Lambda )\psi _{W})(p)=R(\Lambda ,p)\cdot \psi _{W}(\Lambda ^{-1}p)
\end{equation}
The transition to covariant wave function and fields is done with the help
of intertwiners $u(p,s_{3})$ resp. the rectangular matrix $U(p)$ constructed
from their 2s+1 column vectors of length ($2A+1)\cdot (2B+1)$%
\begin{equation}
U(p)D^{(s)}(R())=D^{(A,B)}(\Lambda )U(\Lambda ^{-1}p)
\end{equation}
i.e. within Wigner's Poincar\'{e} group positive energy representation
theory one can intertwine the rotations (with the p-dependent Wigner
R-matrix) with the (dotted and undotted) finite dimensional spinor
representations $D^{(A,B)}.$ Since the $D^{(s)}$ representation of the
rotations is ``pseudo-real'', there exists another intertwiner matrix $V(p)$
which is ``charge-conjugate'' to $U(p).$ To each of the infinitely many
intertwiner systems (the only restriction on A,B for given physical spin s
is $\left| A-B\right| \leq s\leq \left| A+B\right| )$ one has a local field
obeying the spin-statistics connection: 
\begin{equation}
\psi ^{(A,B)}(x)=\frac{1}{\left( 2\pi \right) ^{\frac{3}{2}}}\int \left(
e^{-ipx}\sum u(p,s_{3})a(p,s_{3})+e^{ipx}\sum v(p,s_{3})b^{\ast
}(p,s_{3})\right) \frac{d^{3}p}{2p_{0}}  \label{field}
\end{equation}
where $a,b$ are the (creation) annihilation operators associated with the
Fock space enlargement of the Wigner representation space and hence
independent of the choice of intertwiners. All the different fields are
describing the same $(m,s)$ particle physics and live in the same Fock
space. They constitute only the linear part of a huge (Borchers) equivalence
class of fields. For free fields, this equivalence class contains in
addition all Wick-monomials, and it is well known that they are
indispensible for introducing perturbative interactions. The above different 
$\psi ^{\prime }s$ can be mutually solved: 
\begin{equation}
\psi ^{(A^{\prime },B^{\prime })}(x)=M_{(A,B)}^{(A^{\prime },B^{\prime
})}(\partial )\psi ^{(A,B)}(x)
\end{equation}
where $M_{(A,B)}^{(A^{\prime },B^{\prime })}(\partial )$ is a rectangular
matrix (matrix indices supressed) involving $\partial _{\mu }$ derivatives.

Explicit formulas can be found in the first volume of \cite{Wein}. Among the
infinitely many possibilities essentially only one is ``Lagrangian'' i.e.
can be used in a quantization approach starting from a classical Hamiltonian
principle. The other descriptions are physically equally acceptable, since
there is no quantization principle which enforces to do quantum physics
through a classical parallelism with the Lagrangian formalism. In fact they
describe the same physics in form of a different ``field coordinatisation''.

Indeed for LQP, pointlike fields (\ref{field}) are like coordinates in
differential geometry; it may be sometimes convenient to use them but
structural theorems on charge-carrying fields (classification of statistics,
including braid group statistics for low dimensional charge carriers,
TCP...) and internal symmetries (symmetries and their spontaneous breaking,
the Schwinger-Higgs screening mechanism...) are best done in terms of the
properties of the net: 
\begin{equation}
\mathcal{O}\rightarrow \mathcal{A}(\mathcal{O})
\end{equation}
The causality and spectral properties of these nets constitute the physical
backbone of LQP. The notion ``local'' is then extended to all Boson and
Fermion fields, because they allow an unrestricted iterative application to
the vacuum without encountering local annihilators, and therefore such an
extension preserves the important properties of the original observables.
More general charge carrying fields which extend the above local (bosonic or
fermionic) net are called ``localizable''(with respect to the observables).
In particular plektonic (braid-group statistics) d=1+2 dimensional fields
can never have a Fock space structure and always locally annihilate charge
sectors when the operator domain does not match the range of the charge
sector of the state vector. Although such fields (as some fields used in
gauge theory) have necessarily a semi-infinite (spacelike) string-like
extension, these charge carriers are associated with a local net of
observables i.e. they do not bring in an aspect of elementary length or any
other restriction of the causality principle. A genuinely nonlocal theory
would \textit{violate causality in its observable algebra}; as long as the
theory admits a causal observable algebra there is no elementary length,
independently of the possibly extended nature of charged operators. With
other words extended operators which transfer charges and communicate
between different representations of the observables are permitted as long
as their commutation relations relative to the observables reflect their
spatial extension in the previously mentioned sense.

It is important to note that the Wigner free fields have operator dimensions
(referring to the short distance power behaviour) which increase with spin: $%
dim\psi _{(s=0})=1,\,\,dim\psi _{(s=\frac{1}{2}})=\frac{3}{2},\,\,dim\psi
_{(s=1})\geq 2.$ This is the deeper reason why the incorporation of
interacting theories into the scheme of causal renormalized perturbation
requires special cohomological tricks (BRS) for $s\geq 1$ (the LQP version
of gauge theories, see next section).

The Wigner approach for ($m=0,s\geq 1)$ leads to a more restricted class of
intertwiners, since many representations (e.g. the $D^{(\frac{1}{2},\frac{1}{%
2})}$ vector representation$),$ as a result of the different nature of the
``little group, \textit{cannot} be intertwined with the physical photon
(0.h=1) of the Wigner representation theory. In fact the range of
dotted/undotted indices in \ref{field} is restricted according to $h=\pm
\left| A-B\right| $ \cite{Wein}. There are two methods to overcome this
restriction; one physical way of introducing a semiinfinite spacelike
localized vectorpotential $A_{\mu }(x,n)$ depending on a spacelike string
direction $n$ into the Wigner photon space, or the extension by ghost fields
(indefinite metric or different star-operation) formalism which keeps the
formal Lorentz-covariance (together with the point-like nature) in the form
of ``pseudo-unitarity'' representations). Whereas the first method is
physically deeper and more promising, the second one is the only one which
is compatible with the presently known formalism of renormalized
perturbation theory. The latter does not care whether the locality is formal
instead of physical and whether the boost transformations are pseudo-unitary
instead of unitary, but the interpretation does.

The remaining positive energy representations are Wigner's famous
``continuous spin'' representation which are infinite component (infinite
dimensional representations of the massless ``little group''). They are
usually dismissed by saying that nature does not make use of them. Apart
from the fact that a theoretician should not argue in this way (and in fact
he doesn't if it comes to supersymmetry), the dismissal is probably founded
on the naive identification of irreducible positive energy representation
with physical particles. This ignores that particles should be described by
states, which in addition to forming irreducible positive energy
representations, must also have good localization properties. The modular
localization method below reveals that any positive energy representation
can be localized in wedges. For all positive energy representations with
finite spin/helicity the localization can be sharpened; for the m=0
continuous spin representations however the same methods are inconclusive.
It is doubtful that they admit a sharper localization, needed for particle
interpretation including scattering, and this may cause their
disqualification as candidates for physical particles on the theoretical
side. There are also many useful particle-like objects or states which are
not described by (m,s=semi-integer) Wigner representations as e.g.
infraparticles (electron with photon cloud), ultraparticles, quarks... \cite
{Bu}). The borderline between physical particle and other weakly localizable
objects is the stringlike (more appropriatly spacelike-cone) localization.
This localization is still sufficient to derive scattering theory and on the
other hand it follows from the existence of field theoretic charge sectors
which fulfill the mass gap assumption \cite{Haag}. Operators with braid
group commutation relations in d=1+2 which have one-particle components with
mass gaps, are necessarily stringlike and lead to anyons (abelian, spin
arbitrary) or plektons (nonabelian, spin quantized). Therefore compactly
(e.g. double cone) localizable fields and particles in d=1+2 are only
consistent with the permutation group statistics which is a special case of
braid group statistics.

If fields are analogous to coordinates in differential geometry, there
should be a way to at least construct interaction free nets directly,
without ever using free fields. The idea behind this is to characterize
wedge localized real subspaces in Wigner space with the help of modular
operators (instead of Cauchy initial value data). Assumefor simplicity
integer spin selfconjugate Bosons and define a real subspace $H_{R}(W_{st})$
of $H_{Wigner}$ as: 
\begin{eqnarray}
H_{R}(W_{st}) &=&closure\,\,of\,\,\func{real}\,\,lin.\,comb.\left\{ \psi
\mid s\psi =\psi \right\}  \\
s &\equiv &j\delta ^{\frac{1}{2}},\,\,\,\,\,s^{2}=1  \nonumber
\end{eqnarray}
The notation is as follows: $\delta ^{i\tau }:=U(\Lambda _{x,t}(2\pi \tau ))$
is the Lorentz boost in the x-t direction associated to the standard x-t
wedge $W_{st}:=\left\{ x\in \Bbb{R}^{4};x_{1}>\left| x_{0}\right| \right\} $%
, and $j=\theta \cdot rot_{x}(\chi =\pi )$ is, apart from a $\pi $-rotation
around the x-axis, the antiunitary TCP transformation $\theta $ acting on
the Wigner one-particle space, which for non-selfconjugate particles
consists of a direct sum of the particle and antiparticle space. The
unbounded $\delta ^{\frac{1}{2}}>0$ is defined by functional calculus from $%
\delta ^{it}$ and has a domain consisting of boundary values of analytically
continuable 2s+1 component wave function which have the momentum space
rapidity ($p_{0}=m\cosh \theta ,\,p_{x}=m\sinh \theta $ ) analyticity in the
strip $-\pi <Im\theta <0.$ $s$ inherits the densely defined domain from $%
\delta ^{\frac{1}{2}}$ and the antilinearity from $j.$ The best way to
describe this real Hilbert space of wedge localized functions is to say that
they are strip-analytic and fulfill Schwartz reflection principle around the
line $Imz=-\frac{i\pi }{2}.$ In case of antiparticles$\neq $particles one
most double the number of components and use the full charge conjugated wave
functions in the reflection principle instead of just the complex conjugate.
This is closely related to the crossing ``symmetry'' (it is not a symmetry
in the standard operational sense of QT) in interacting systems (see later).
The involutive property $s^{2}=1$ on this domain, in mathematical notation $%
s^{2}\subset 1$, is a consequence of this definition. Such unbounded (but
yet involutive) operators did not occur in any other area of mathematical
physics and therefore are not treated in books on mathematical methods. In
fact they seem to be characteristic of the Tomita-Takesaki modular theory.
It is precisely the combination of unboundedness and involutiveness which is
responsible for the emergence of localization and geometrical properties
from domain properties of quantum physical operators. The real closed
subspace may be used to define a dense wedge localization space\footnote{%
A change of sign in the definition of $H_{R}(W)$ would not change the dense
complex localization space (which is a Hibert space in the graph $s$-norm).} 
$H(W_{st})\equiv $ $H_{R}(W_{st})+iH_{R}(W_{st})$ on which the operator $s$
acts as: 
\begin{equation}
s(h+ih)=h-ih
\end{equation}
$H_{R}(W_{st})$ is ``standard'' i.e. 
\begin{eqnarray}
&&H_{R}(W_{st})\cap iH_{R}(W_{st})=\left\{ 0\right\}  \\
&&H(W_{st})\equiv
H_{R}(W_{st})+iH_{R}(W_{st})\,\,is\,\,dense\,\,in\,\,H_{Wigner}  \nonumber
\end{eqnarray}
The natural localization topology is the graph norm of $s$. It is somewhat
unusual and treacherous that the formula for $s$ looks so universal and the
differences in the localization for different wedges $W_{\Lambda }:=\Lambda
W_{st}\,,\,W_{a}:=T(a)W_{st}$ is solely encoded in the domain of definition
of $s_{(\Lambda ,a)\text{ }}($i.e. only where and not how it acts) which he
usually considers to be a fine and somewhat irrelevant technical point. For
positive energy representations the geometric inclusion $W_{a}:=T(a)W_{st}%
\subset W_{st},\,a\in W_{st}$ $($translating wedges into themselves) implies
the proper inclusion (D. Guido, private communication 1996) $%
H_{R}(W_{a})\subset H_{R}(W_{st}),$ in fact the geometric inclusion
properties are equivalent to the positive spectrum condition. For the
understanding of the latter claim one has to decompose the spacelike $a$
into two lightlike components $a_{\pm }$ for which one takes of course the
two lightlike vectors by which the wedge $W_{st}$ is generated. Different
from spacelike translations, these lightlike translations have a positive
generator.

Having constructed a net of wedge-localized real subspaces $H_{R}(W),$ one
may move ahead and introduce compactly localized spaces $H_{R}(\mathcal{O})$
through intersections $\cap _{W\supset \mathcal{O}}W$ 
\begin{equation}
H_{R}(\mathcal{O})=\cap _{W\supset \mathcal{O}}H_{R}(W)  \label{inters}
\end{equation}
In order to insure the nontriviality of these intersections, one needs to
restrict the positive energy representations to those with a
finite-dimensional representation of the Wigner ``little group'' which
amounts to (half)integer spin/helicity. In this way one obtains e.g. the net
of double cones; a direct construction of the associated modular objects is
more difficult because the modular group behaves ``geometric'' (i.e. as a
diffeomorphism of Minkowski space) only asymptotically close to the
``horizon'' (the boundary of the causal closure) of the region. Note that in
order to define these localization spaces, we did not use any $u,v$
intertwiners. If we had done this, the present intrinsic concept of
localization would have been lost and we would have been back at x-space
properties of covariant wave functions or pointlike fields i.e. those field
coordinatisations which destroyed the unicity. The size of localization is
contained in certain Payley-Wiener type of bounds in imaginary momentum or
rapidity directions.

The last step to the nets consists (say for the case of integer spin) in the
application of the Weyl functor which maps real subspaces into the von
Neumann subalgebras of a net: 
\begin{eqnarray}
&&H_{R}(\mathcal{O})\stackrel{\mathcal{F}}{\rightarrow }\mathcal{A}(\mathcal{%
O}) \\
\mathcal{A}(\mathcal{O}) &=&\limfunc{alg}\left\{ W(f)\mid f\in H_{R}(%
\mathcal{O})\right\}   \nonumber \\
W(f) &=&e^{i(a^{\ast }(f_{1})+h.c.)+i(b^{\ast }(f_{2})+h.c.)},\,\,\,f=\left(
f_{1},f_{2}\right)   \nonumber
\end{eqnarray}
where $b^{\#},a^{\#}$ stand for (anti)particle Wigner creation and
annihilation operators. The functor $\mathcal{F}$ is orthocomplemented i.e.
the symplectic or (by multiplication with $i)$ real orthogonal complement of
a real subspace is mapped into the von Neumann algebraic commutant. The
images $J,\Delta ^{it},S$ of $j,\delta ^{it},s$ under $\mathcal{F}$ are the
modular objects of the algebraic version of the Tomita Takesaki modular
theory for the special case of the pair ($A(W_{st}),\Omega )$ of wedge
algebra and vacuum vector\footnote{%
A construction of the free net without using modular localization methods
can be found in \cite{Baumgärtel}. It is however the modular method which
extends to the interacting case.}.

The general theory says that for a von Neumann algebra $\mathcal{A}$ with a
cyclic and separating vector $\Omega $ , the definition: 
\begin{equation}
SA\Omega =A^{\ast }\Omega ,\,A\in \mathcal{A}
\end{equation}
introduces a closable operator, whose polar decomposition; 
\begin{equation}
S=J\Delta ^{\frac{1}{2}}
\end{equation}
defines a unitary $\Delta ^{it}$ and a antiunitary involution $J$ which are
of fundamental significance for the pair ($\mathcal{A},\Omega ).$ The
operator $\Delta ^{it}$ defines the ``modular'' automorphism $\sigma _{t}$
of $\mathcal{A}$ (a kind of generalized hamiltonian) with respect to $\Omega 
$ and $J$ the modular involution $j$ (a kind of generalized TCP reflection): 
\begin{eqnarray}
\sigma _{t}(\mathcal{A}) &=&\mathcal{A},\,\,\,\,\sigma _{t}(A)\equiv \Delta
^{it}A\Delta ^{-it} \\
j(\mathcal{A}) &=&\mathcal{A}^{\prime },\,\,\,\,j(A)\equiv JAJ  \nonumber
\end{eqnarray}
This basic theorem was stated and proved by Tomita with significant
improvements due to Takesaki \cite{Bra Ro}. In the context of thermal
quantum physics it received an important independent contribution in form of
the KMS condition from Haag Hugenholz and Winnink; whereas Kubo, Martin and
Schwinger only used this analytic condition in order to avoid the
calculation of traces, the HHW paper elevates this property to one of the
most important conceptual tools related to stability of states and to the
second thermodynamical law \cite{Haag}. Its relevance for localization in
QFT was first seen in full generality by Bisognano and Wichmann \cite{Bi
Wich} and the thermal aspects of (wedge) localization (the Hawking-Unruh
connection) were first stressed by Sewell \cite{Sew}. Although we explained
the construction of free nets only for Bosons, the formalism adapts easily
to Fermions. Fermions are preempted in the modular localization of the
Wigner theory by the appearance of a mismatch between the geometrical
opposite of $H_{R}(W)$ obtained by a 180 degree rotation, and its symplectic
or real orthogonal complement. This leads to a modification of the Tomita
involution in form of an additional twist which can be shown to preempt the
Fermi-statistics. Our inverse use of the Bisognano-Wichmann idea for the
purpose of direct net construction which we exemplified for free theories in
arbitrary spacetime dimensions can be generalized to interacting theories
with the mathematical control being restricted presently to d=1+1. Some of
these results will be presented in the last section.

Already in the very early development of algebraic QFT \cite{Haag-Sch} the
nature of the single local von Neumann algebras became an interesting issue.
Although it was fairly easy (and expected) to see that i.e. wedge- or double
cone- localized algebras are von Neumann factors (in analogy to the tensor
product factorization of standard QT under formation of subsystems, it took
the ingenuity of Araki to realize that these factors were of type $III$
(more precisely hyperfinite type $III_{1}$ as we know nowadays, thanks to
the profound contributions of Connes and Haagerup), at that time still an
exotic mathematical structure. Hyperfiniteness was expected from a physical
point of view, since approximatability as limits of finite systems (matrix
algebras) harmonizes very well with the idea of thermodynamic+scaling limits
of lattice approximations. A surprise was the type $III_{1}$ nature which,as
already mentioned, implies the absence of pure states (in fact all
projectors are Murray von Neumann equivalent to the identity operator) on
such algebras; this property in some way anticipated the thermal aspect
(Hawking-Unruh) of localization. Overlooking this fact which makes local
algebras significantly different from algebraic aspects of QM, it is easy to
make conceptual mistakes which could e.g. suggest an apparent breakdown of
causal propagation. For the discussion of such a kind of error and its
correction see \cite{Yng-Buch}), as already mentioned in the introduction.
If one simply grafts concepts of QM onto the causality structure of LQP
(e.g.quantum mechanical tunnelling, structure of states) without deriving
them in LQP , one runs the risk of wrong conclusions about e.g. the
possibility of superluminal velocities.

Let me, at the end of this section mention two more structural properties,
intimately linked to causality, which distinguish LQP rather sharply from
QM. One is the Reeh-Schlieder property: 
\begin{eqnarray}
\overline{\mathcal{P}(\mathcal{O})\Omega } &=&H,\,\,cyclicity\,\,of\,\,\Omega
\label{cyc} \\
A\in \mathcal{P}(\mathcal{O}),\,\,A\Omega &=&0\Longrightarrow
A=0\,\,i.e.\,\,\Omega \,\,separating  \nonumber
\end{eqnarray}
which either holds for the polynomial algebras of fields or for operator
algebras $\mathcal{A}(\mathcal{O}).$ The first property, namely the
denseness of states created from the vacuum by operators from arbitrarily
small localization regions (a state describing a particle behind the moon%
\footnote{%
This weird aspect should not be held against QFT but rather be taken as
indicating that localization by a piece of hardware in a laboratory is also
limited by an arbitrary large but finite energy, i.e. is a ``phase space
localization'' (see subsequent discussion). In QM one obtains genuine
localized subspaces without energy limitations.} and an antiparticle on the
earth can be approximated inside a laboratory of arbitrary small size and
duration) is totally unexpected from the global viewpoint of general QT. In
the algebraic $\mathcal{A}(\mathcal{O})$ formulation this can be shown to be
dual to the second one (in the sense of passing to the commutant), in which
case the cyclicity passes to the separating property of $\Omega $ with
respect to $\mathcal{A}(\mathcal{O}^{\prime }).$

Of course the claim that somebody causally separated from us may provide us
with a dense set of states is somewhat unusual if one thinks of the
factorization properties of ordinary QT. The large enough commutant required
by the latter property is guarantied by causality (the existence of a
nontrivial $\mathcal{O}^{\prime })$ and shows that causality is again
responsible for the unexpected property. If the naive interpretation of
cyclicity/separability in the Reeh-Schlieder theorem leaves us with a
feeling of science fiction (and also has attracted a lot of attention in
philosophical quarters), the challenge for a theoretical physicist is find
an  argument why, for all practical purposes, the situation nevertheless
remains similar to QM. This amounts to the fruitful question: which among
the dense set of localized states can be really produced with a controllable
expenditure (of energy)? In QM the asking of this question is not necessary,
since the localization at a given time via support properties of wave
functions leads to a tensor product factorization of inside/outside so that
the inside state vectors are automatically never dense in the whole space.
Later we will see that most of the very important physical and geometrical
informations are encoded into features of dense domains, in fact the
aforementioned modular theory is explaining such relations. For the case at
hand the reconciliation of the paradoxical aspect of the Reeh-Schlieder
theorem with common sense has led to the discovery of the physical relevance
of \textit{localization with respect to phase space in LQP}, i.e. the
understanding of the \textit{size of degrees of freedom} in the set: 
\begin{eqnarray}
P_{E}\mathcal{A}(\mathcal{O})\Omega \,\,is\;compact &&\,\,\, \\
\,\,e^{-\beta \mathbf{H}}\mathcal{A}(\mathcal{O})\Omega \;is\,\,nuclear,\,\,%
\mathbf{H} &=&\int EdP_{E}\,\,  \nonumber
\end{eqnarray}
The first property was introduces way back by Haag and Swieca \cite{Haag}
whereas the second statement (and similar nuclearity statements involving
modular operators of local regions instead of the global hamiltonian) which
is more informative and easier to use, is a later result of Buchholz and
Wichmann \cite{Bu-Wich}. It should be emphasized that the LQP degrees of
freedom counting of Haag-Swieca, which gives an infinte but still compact
set of localized states is different from the finiteness of degrees of
freedom per phase space volume in QM, a fact often overlooked in present
day's string theoretic degree of freedom counting. The difference to the
case of QM disappears if one uses instead of a strict energy cutoff a Gibbs
damping factor $\ e^{-\beta H}$ as above$.$ In this case the map $\mathcal{A}%
(\mathcal{O})\rightarrow e^{-\beta H}\mathcal{A}(\mathcal{O})\Omega $ is
``nuclear'' if the degrees of freedom are not too much accumulative (which
then would cause the existence of a maximal Hagedorn temperature. The
nuclearity assures that a QFT, which was given in terms of its vacuum
representation, also exists in a thermal state. An associated nuclearity
index turns out to be the counterpart of the quantum mechanical Gibbs
partition function \cite{Haag} and behaves in an entirely analogous way.

The peculiarities of the above Haag-Swieca degrees of freedom counting are
very much related to one of the oldest ``exotic'' and at the same time
characteristic aspects of QFT namely vacuum polarization. As discovered by
Heisenberg, the partial charge: 
\begin{equation}
Q_{V}=\int_{V}j_{0}(x)d^{3}x=\infty 
\end{equation}
diverges as a result of uncontrolled vacuum fluctuations near the boundary.
For the free field current it is easy to see that a better definition
involving test functions, which takes into account the fact that the current
is a 4-dim distribution and has no restriction to equal times, leads to a
finite expression. The algebraic counterpart is the so called ``split
property'', namely the statement \cite{Haag} that if one leaves between say
the double cone (the inside of a ``relativistic box'') observable algebra $%
\mathcal{A}(\mathcal{O})\,$and its causal disjoint (its relativistic
outside) $\mathcal{A}(\mathcal{O}^{\prime })$ a ``collar'' $\mathcal{O}%
_{1}\cap \mathcal{O}$, i.e. 
\begin{equation}
\mathcal{A}(\mathcal{O})\subset \mathcal{A}(\mathcal{O}_{1}),\,\,\,\mathcal{%
O\ll O}_{1}\,,\,\,properly
\end{equation}
then it is possible to construct in a canonical way a type $I$ tensor factor 
$\mathcal{N}$ which extends into the collar $\mathcal{A}(\mathcal{O}%
)^{\prime }\cap \mathcal{A}(\mathcal{O}_{1})$ i.e. $\mathcal{A}(\mathcal{O}%
)\subset \mathcal{N}\subset \mathcal{A}(\mathcal{O}_{1}).$ With respect to $%
\mathcal{N}$ the vacuum state factorizes i.e. as in QM there are no vacuum
fluctuations for the ``smoothened'' operators in $N.$ The algebraic analogon
of Heisenberg's smoothening of the boundary is the construction of a
factorization of the vacuum with respect to a suitably constructed type $I$
factor algebra which uses the collar extension of $\mathcal{A}(\mathcal{O}).$
It turns out that there is a canonical, mathematically distinguished
factorization, which lends itself to define a natural ``localizing map'' $%
\Phi $ which has given valuable insight into an intrinsic LQP version of
Noether's theorem \cite{Haag}, i.e. one which does not rely on any
parallelism to classical structures as is the case with quantization. It is
this ``split inclusion'' which allows to bring back the familiar structure
of QM since type I factors allow for pure states, tensor product
factorization, entanglement and all the other properties at the heart of
quantum theory and the measurement process.

There are also interesting ``folklore theorems'' i.e. statements which are
mostly taken for granted, but for which yet no rigorous argument exists (but
also no counter-example). One is the statement of ``nuclear democracy''. In
the context of LQP it states that an operator from a (without loss of
generality) double cone algebra $A\in \mathcal{A}(\mathcal{O})\,$or a
pointlike field couples to all states to which the superselection rules
allow a nonvanishing matrixelement. In particular we expect: 
\begin{equation}
\left\langle \varphi ^{in}\left| A\right| \psi ^{in}\right\rangle \neq 0
\end{equation}
if the (say incoming) multiparticle state vector $\varphi ^{in}$ lies in the
same charge superselection sector as $A\left| \psi ^{in}\right\rangle $,
i.e. ``everything communicates with everything'' as long as the charges match%
\footnote{%
This forces the substitution of the QM hierarchical concept of bound state
particles in favor of charge fusion in LQP, which in turn means ``nuclear
democracy'' between particles.}. A special case is the phenomenon of vacuum
or better one-particle polarization through interaction i.e. the idea that
there may be no interacting local operator $A\in \mathcal{A}(\mathcal{O})\,$%
at all such that $A\Omega $ is in the one-particle space without additional $%
p\bar{p}$-contributions. In order to suppress this p\={p} polarization cloud
in state vectors of interacting theories, one has to allow at least a
semiinfinite localization region as the wedge region. For any compact region
or even noncompact regions which are tiy bit smaller than wedges, the
infinite particle clouds and the field point of view take over. The
polarization cloud content of a state vector $A\Omega $ with $A\in \mathcal{A%
}(\mathcal{O})$ is intimately related to the modular objects of $(\mathcal{A}%
(\mathcal{O}),\Omega ).$ If one could back up these expectations (based on
model observations) by rigorous theorems, one would have achieved an
intrinsic understanding of interactions. The section 5 gives a brief account
on what is presently known about modular construction of interacting nets.

\section{{\protect\large Renormalized Perturbation, Problems with s}$\geq 1$}

Following Tomonaga, Feynman and Schwinger and the other pioneers of
perturbative renormalization, interactions are traditionally introduced
through one of the various forms of quantization (canonical, path
integral,..).

The method which brings out the pivotal role of causality in the most
explicite way is however the so called ``causal perturbation method'' of
Stuekelnberg and Bogoliubov \cite{Bog} which was formulated as a finite
iteration method within the principles of LQP without reference to
quantization by Epstein and Glaser \cite{E-G}. Some refinements of that
method, notably related to curved space time and gauge theories, have been
added recently by \cite{Brun Fred}\cite{D-F}. Also Weinberg's more formal
derivation of Feynman rules for arbitrary spin \cite{Wein} is somewhat in
the spirit of causal perturbations.

It is a conceptual weakness of any quantization approach that contrary to
QM, where this can be given a rigorous meaning, quantization in field theory
remains more on the intuitive artistic side. Only for a so-called
superrenormalizable interactions is the assumed canonical or functional
Feynman-Kac quantization structure also reflected in the renormalized
result; in all other cases it only serves as a vehicle which activates
physicists thought and does not survive the renormalization procedure: i.e.
with the mentioned exception no renormalized result fulfills canonical
commutation relations or functional integral representations, rather the
only surviving structure is causality/locality. This artistic rather than
mathematical aspect pervades the standard text book formulation of QFT. Such
a state of affairs is acceptable, as long as one remains aware that (what I
will summarily call) the Lagrangian quantization is basically an efficient
chain of formal manipulations and tricks which lead from slightly wrong
assumptions after some repair to the correct perturbative results. Wheras
the canonical structure and the functional integral representation cannot be
upheld, the physical causality properties do survive the necessary repair
procedure better known under the name of renormalization.

In order to rescue the canonical or functional structures at any costs,
physicist sometimes resort to imagine the existence of physical cutoffs or
regulators and use the euphemism ``cutoff canonical variables or cutoff
functional representations'' without confronting those problems of
noncausal/nonlocal theories mentioned in the introduction. In this way of
thinking, the infinities of the unrenormalized theory relative to the
renormalized are attributed sometimes more physical significance than just
indicating the necessity of repairing a slightly incorrect classical
starting point (the classical Poincar\'{e}-Lorentz- instead of the
Wigner-particle picture), which would be avoided in the causal perturbative
approach.

To be fair, these conceptual drawbacks of the quantization artistry are
partially offset by the efficiency of renormalizing away infinities through
Feynman rules. Even if e.g. Schwinger's finite split point method for the
nonlinear terms in field equations may be conceptually cleaner because one
never meets an infinity (as long as one does not interchange short distance
limits with the other operations), but is practically less efficient as
Feynman's infinity (or ad hoc cutoff) method.

Different from quantization +repair of infinities, LQP only uses those
physical assumptions which are also genuinely reflected in the results
(causality, spectral properties, modular structure of local algebras,..).
The principles are the same principles as standard QFT but it does so in a
more conscientious way. In such an approach the short distance properties of
individual fields are, apart from perturbation theory (infinitesimal
deformations around free fields), less tightly connected with the existence
of the model. We will come back to this important point in the
nonperturbative section 5. In the following we will illustrate the strength
of the LQP point of view in perturbation theory. The renormalized results
are of course the same as in the functional approach, but the derivation and
the guiding physical ideas differ in an interesting way.

In causal perturbation theory, which may be considered as a particular form
of perturbative LQP, the interaction is implemented by locally coupling the
free fields (any choice possible, $\psi $ does not have to be Lagrangian!)
by an L-invariant sum over Wick monomials $W_{i}(x)$ and one defines the
following formal transition operator in Fock space\footnote{%
There is no compelling physical reason besides the historical success in QED
and the analogy with QM why outside of deformation of free fields the
introduction of interactions should follow this pattern. The existence of
perturbation theory in the sense of a deformation theory has in general no
bearing on the existence of an associated nonpertubative version.}: 
\begin{eqnarray}
S(g,h) &=&Te^{i\int \left\{ g(x)W(x)+h(x)\psi (x)\right\} d^{4}x}
\label{Bogol} \\
\tilde{C} &\subset &\limfunc{supp}g\subset C  \nonumber \\
g_{i} &=&const\,\,in\,\,\tilde{C}  \nonumber
\end{eqnarray}
where $W(x)=\sum g_{i}W_{i}(x)$ and $C,\tilde{C}$ are large double cone
regions. In the following we specialize to one field and one coupling for
simplicity of notation (the notation for the general case with several
fields and monomials we leave to the reader). Already without the
time-ordering $T$, the operator exponential is a mathematically delicate
object since the smeared Wick-powers beyond the second are not essentially
selfadjoint on their natural domains. With the time ordering it is more
serious: apart from certain $W$'s with low operator dimensions (a situation
which cannot occur in d=1+3 dimensions), there is no operator functional $%
S(g)$ in Fock space for which a mathematical control has been achieved (no
solution of the ``Bogoliubov axiomatics'' in d=1+3). Causal perturbation
theory does not attempt to make sense of $S(g)$ but only of its $n^{th}$
order power series term in g. Therefore one proceeds along the following two
lines:

\begin{itemize}
\item  \textbf{Extraction of general causality properties for }$S(g)$\textbf{%
\ and related operators (the ``Bogoliubov axiomatics'').} The basic
causality in the time-ordered formalism is: 
\begin{eqnarray}
T(\psi (x_{1})...\psi (x_{n})) &=&T(\psi (x_{1}))..\psi (x_{k}))\cdot T(\psi
(x_{k+1})..\psi (x_{n})) \\
if\,\,x_{j} &\notin &x_{i}+\bar{V}_{+},\,\,i=1,...,k,\,\,j=k+1,...,n 
\nonumber
\end{eqnarray}
For the purpose of (formally) extracting a causal net it is helpful to
reformulate this property in terms of another relative transition operator: 
\begin{eqnarray}
V(g,h) &\equiv &S(g,h=0)^{-1}S(g,h) \\
causality &:&\,\,V(g,h_{1}+h_{2})=V(g,h_{1})V(g,h_{2})  \nonumber \\
if\,\,\limfunc{supp}h_{1} &\notin &\limfunc{supp}h_{2}+\bar{V}_{+}  \nonumber
\end{eqnarray}
With the local algebras being now defined as (the notation $\func{alg}$
includes the von Neumann closure): 
\begin{equation}
\mathcal{A}_{g}(\mathcal{O})\equiv \func{alg}\left\{ V(g,h),\limfunc{supp}%
h\subset \mathcal{O}\right\}
\end{equation}
In fact a change of the coupling strength g outside $C$ (see \ref{Bogol})
does not change the net $\mathcal{A}_{g}(\mathcal{O})$ for $\mathcal{O}$
inside $\tilde{C},$ except for a common unitary (the nets are isomorphic
i.e. considered to be identical) 
\begin{eqnarray}
&&V(g+\delta g,h)=AdU(g,\delta g)V(g,h) \\
&&\limfunc{supp}\delta g\text{\thinspace \thinspace \thinspace }%
outside\,\,\,\,\tilde{C}  \nonumber
\end{eqnarray}
With this formula, the transition from the BPS-EG to the LQP net formalism
has been achieved \cite{Brun Fred}. The algebraic content has been
constructed in an auxiliary Fock space whose particle content is not
necessarily identical with the physical particle content and the adiabatic
limit of the E-G approach (which would have forced the coalescence of the
two) has been avoided.

\item  \textbf{Perturbation as a deformation of free fields.} Having no
control over the objects in the Bogoliubov axiomatics, we satisfy ourselves
with existence and properties of causal power series for $S(g):=S(g,h)\mid
_{h=0}$ 
\begin{equation}
S(g)=\sum \frac{i^{n}}{n!}\int g(x_{1})....g(x_{n})TW(x_{1})....W(x_{n})
\end{equation}
which allows an iterative construction in $n$ with $W$ serving as the input.
The main inductive step is the construction of the total diagonal part in
n+1 order, assuming that the $n^{th}$ order time ordered product has been
fully (i.e. as an operator-valued distribution on all Schwarz test
functions) constructed. Causality defines the n+1 order object on all test
functions which vanish on totally coalescent diagonal point \cite{Brun Fred}%
. The (Hahn-Banach) extension problem allows for totally locally supported
terms with a priori undetermined coefficient. These local terms are often
(as ``counter-terms'') lumped together with the n=1 term. Mere perturbative
locality and unitarity requirements do not fix this ambiguity (i.e.
perturbatively one always operators in Hilbert space\footnote{%
This is not necessarily so in other (e.g. functional integral) formulations,
where the connection with operator aspects of QT may get lost (even the
introduction of cut-offs or regularizations is no assurance for maintaining
it).}). Rather the introduction of a suitable degree function allows to
control this ambiguities in terms of a finite number of physical parameters,
at least in the case of so-called renormalizable interactions $W$ with dim$%
W\leq 4=d.$ Perturbation is a deformation around known theories which in the
present case are free fields. It only explores an infinitesimal neighborhood
around free fields and not suited for deciding questions about the
mathematical existence. In fact beyond deformation theory it is not
physically compelling to implement the idea of interactions by coupling free
fields to $W^{\prime }s$ in Fock space. Rather this is the perturbative way
of introducing interactions and not a general consequence of the general
framework. Indeed the nonperturbative attempts based on modular theory use a
different implementation of ``interaction'', as will be shown later. The
causal perturbation theory leads to the same renormalized correlation
functions as e. g. the one based on functional integrals. However, as shown
in the sequel, the physical concepts and calculational rules are somewhat
different. In particular all differential identities (as equations of
motion) can be used freely in the ``on shell'' causal formulation, whereas
this is not the case in the off shell functional (euclidean) approach. For
the (m,s) free fields one may take any of the many possibilities in (\ref
{field}) independent of whether the field results from a classical
Lagrangian (in which case its covariant transformation follows from the
Euler equation of motions) or not. But since for given (m,s) there always
exists a Lagrangian ``field coordinatization'' in terms of which one may
rewrite the given interaction $W,$ one does not loose anything if one starts
from Lagrangians. The main benefit of the causal perturbation viewpoint lies
in the fact that one liberates oneself from the moral obligation to repair
something which came by quantization from classical theory. Instead the main
question is how, by using the terms in the formal power series expansion,
can I obtain something which is well defined in Fockspace, fulfills
causality and unitarity requirements and has the right to be called
time-ordered product of (the well-defined) $W^{\prime }s?$ The last
statement can be made more precise by saying it should coalesce with the
naive time-ordered product of $W^{\prime }s$ if one smeares them with test
functions which have non-coalescent supports. So renormalization in the
causal approach just amounts to an extension of operator-valued
distributions from the subspace of test functions with this restriction to
all test funtions. In addition one has to reparametrize the theory in terms
of physical masses, charges and couplings and use a field normalization
which harmonizes with the asymptotic scattering interpretation. Since there
was no classical (bare) particle picture from quantization, there is nothing
to be repaired by dumping infinities, hence the causal approach is finite as
was the Schwinger point-split methods, albeit much easier to handle than the
latter. For dim$W\leq 4$ the procedure works in terms of obtaining a
deformation theory with finitely many masses, charges and coupling
parameters. To prove that this extension idea works in an inductive manner
is not easy and the explanation of the necessary technical steps would throw
this conceptually oriented presentation out of balance.
\end{itemize}

The above formal counting argument, if taken serious as a definition of
renormalizability, would rule out all massive higher spin $s\geq 1$ fields
as candidates to be used for interaction polynomials $W$ since there are no
intertwiners from the Wigner particle to covariant local representations $%
\psi $ with dim$\psi <2$. For example a massive $s=1$ object in the
vectormeson description has operator dimension $dimA_{\mu }=2$ (the use of
different intertwiners does not improve this increase of quantum versus
classical dimension), so that any trilinear interaction involving $A_{\mu }$
(and lower spin) has $dimW\geq 5.$ Fortunately this barrier against
renormalizability created by Wick-polynomials of free fields involving $%
s\geq 1$ has an interesting loophole, namely it can be undermined by a
``cohomological trick'' which consists in the following observation. One is
asked to find a cohomological representation of the e.g. $(m,s=1)$ physical
Wigner space: 
\begin{equation}
H_{Wigner}=\frac{\ker s}{\limfunc{im}s},\,\,\,s^{2}=0
\end{equation}
Here $s$ acts on $H_{ext}$ and the Poincar\'{e} group is still covariantly
represented on $H_{ext}$ (the pseudo-unitary nature of the boost
representors however turns out to be unavoidable). The transversality of the
covariant inner product of the vectorpotential (which was the origin of dim$%
A_{\mu }=2$ instead of the classical dimension 1) only emerges in the
cohomological descend from $H_{ext}$ to $H_{Wigner}.$ The question why a 
\textit{cohomological} extension and not another one which reduces the
dimension to the classical value, lies in the expectation that cohomological
structures tend to be stable under perturbative deformations i.e. one has
the best chance to return to the cohomology space at the end of the
perturbative calculations. The simplest cohomological extension of the
Wigner wave function space which allows a nilpotent operation $s$ with $%
s^{2}=0,$ such that the physical transversality condition $p^{\mu }A_{\mu
}(p)=0$ follows from the application of $s,$ needs besides two scalar ghosts
wave functions $\omega $ and $\bar{\omega}$ another scalar ghost field $%
\varphi $ (often called the Stueckelnberg field):

\begin{eqnarray}
(sA_{\mu })(p) &=&p_{\mu }\omega (p) \\
(s\omega )(p) &=&0  \nonumber \\
(s\bar{\omega})(p) &=&p^{\mu }A_{\mu }(p)-im\varphi (p)  \nonumber \\
(s\varphi )(p) &=&-im\omega (p)  \nonumber
\end{eqnarray}
One immediately realizes that $s^{2}=0$ and that $s(\cdot )=0$ enforces the
vanishing of $\omega $ and relates $\varphi $ to $p^{\mu }A_{\mu }.$ At this
point there is no grading in the formalism, i.e. the $\omega $ and $\varphi $
are simply ungraded wave functions. However the functorial transition from
Wigner theory to QFT requires the introduction of a grading with $\deg
\omega =1,\deg \bar{\omega}=-1,$ and $\deg A_{\mu }=0,$ with $s$
transferring degree 1. The reason is that only with this grading assignment 
\cite{Dubois} the $s$ allows a natural tensor extension to multiparticle
spaces with stable nilpotency, 
\begin{equation}
s(a\otimes b)=sa\otimes b+(-1)^{\deg a}a\otimes sb
\end{equation}
which insures the commutativity of the Wigner/Fock cohomological ascend and
descend: 
\begin{eqnarray}
&&\stackunder{\downarrow }{H_{ext}}\rightarrow \stackunder{\downarrow }{%
\mathcal{H}_{ext}} \\
&&H_{Wig}\rightarrow \mathcal{H}  \nonumber
\end{eqnarray}
where the calligraphic notation stands for the bosonic Fockspace and its
graded extension.

This suggests to view the Fock space version $\delta $ of $s$ as the image
of a (pseudo) Weyl functor $\Gamma $ as $\delta =\Gamma (s)$ and to write
the $\delta $ in the spirit of a formal Noether symmetry charge $Q$ 
\begin{equation}
Q=\int (\partial _{\mu }A^{\mu }(x)+m_{a}\phi (x))\overleftrightarrow{%
\partial }_{0}\omega (x)d^{3}x=Q^{\dagger }
\end{equation}
The experienced reader will easily recognize that we arived at a special
version of the BRS formalism \cite{BRS} which which remains unchanged by
interactions \cite{K O}.

The Fock space version of $s$ yields an object $\delta $ of a differential
algebra with $\delta ^{2}=0$ which changes the $Z$-grading by one unit and
acts on vectors and operators in $H_{ext}$ similar to a global Noether
charge 
\begin{eqnarray}
\delta A=i\left[ Q,A\right] =\delta A\equiv i\left\{ QA-(-1)^{\deg
A}AQ\right\} && \\
Q\,\,in\,\,H_{ext},\,\,\,\,Q^{2}=0 &&  \nonumber
\end{eqnarray}
Note that the nilpotency together with the formal hermiticity $Q=Q^{\dagger
} $prevents a positive inner product in *-representation of such algebras.
It is costumary (and helpful for mathematical controll) to work with two
inner products, one positive definite in order to stay with the mathematics
of operators in Hilbert spaces, and a Krein operator $\eta $ which is used
to define another indefinite one as well as (pseudo)hermiticity. For many
operators the two notions coalesce (they commute with $Q)$ e.g. for all
Poicar\'{e} generators except Lorentz-boosts. In order to introduce
interactions one now uses the extended formalism in the same way as at the
beginning of this section. For an interaction between vectormesons (for
simplicity without additional matter) one may start with a trilinear
expression ($f_{abc}$ are independent couplings) 
\begin{equation}
W^{A}=f_{abc}:A_{a\mu }A_{b\nu }\partial ^{\nu }A_{c}^{\mu }:  \label{coup}
\end{equation}
which in the extended space has dim$W=4.$ The important question to be
answered now is: what is the criterion which selects the physical operators
in $\mathcal{H}$ \ in every order of perturbation theory ? Obviously they
should commute with $Q$ or rather the physical projection of the commutator
should vanish. In addition to finding local operators with this property,
one is interested in the S-matrix for the scattering of the massive
particles which is the adiabatic limit of $S(g)$ for $g(x)\equiv $ $%
const.=g. $ A sufficient condition on the operator-valued functional $S(g)$
which guaranties this property is that $S(g)$ commutes with $Q$ up to
surface terms in g which are localized in the collar (\ref{Bogol}). For the
W and their time-ordered products which appear as integrands in these
relations this means the validity of the following divergence equations: 
\begin{eqnarray}
\left[ Q,W(x)\right] &=&i\partial _{\mu }^{x}W_{1}^{\mu }(x) \\
\left[ Q,T(W(x_{1})...W(x_{n}))\right] &=&i\sum_{l=1}^{n}\partial _{\mu
}^{x_{l}}T(W(x_{1})...W_{1}^{\mu }(x_{l})...W(x_{n}))  \nonumber
\end{eqnarray}
The $W_{1}$ must be constructed in the process of checking these relations.
These equations where introduces by \cite{Scharf} and called ``operator
gauge invariance''. Whereas we will use these divergence relation, we will
not follow this terminology because it creates the erronous impression that
a QFT involving massive vectormesons has to rely on a gauge principle in
addition to renormalizability and the cohomological return to physics. It
turns out that the to the contrary of what happens with low spin s%
\TEXTsymbol{<}1, the renormalization+cohomological descend requirement (the
latter having no counterpart for low spin) are in fact so stronly
restrictive, that not only the masses ase forced to be equal and the
couplings in (\ref{coup}) have to fulfill the Jacobi identity known from
Lie-algebra structure, but all other couplings (including the quadrilinear
couplings induced from the divergence equations) are such that modulo
renormalization terms, they follow the pattern of classical gauge group
theory even though the group theory is not required by physical symmetries.
However the relation to the differential-geometric gauge structure is the
opposite from that in the standard literature. Whereas classical gauge
principles, which selects among the many polynomial couplings (increasing
number with increasing spin) involving vector fields those which nature
(classical e.m.) prefers, usually enter QFT via quantization, the LQP
approach produces a unique interaction between massive vectormesons in the
way sketched before. In particular one obtains the inverse of the 't Hooft
renormalization statement namely the zero mass \textit{(semi)classical limit
of the unique perturbatively renormalizable massive vectormeson theory is a
classical gauge theory}. Without going into more details \cite{D-Schr} we
will collect the important results of the above causal perturbation approach

\begin{itemize}
\item  The masses of the vectormesons are equal and the coupling among
vectormesons and ghosts is determined by one coupling strength. The theory
would show inconsistencies in higher than first order without the
introduction of additional \textit{physical} degrees of freedom. The minimal
(and perhaps only) possibility are (Higgs) scalars but without the usual
vacuum expectations which go with the name of ``Higgs mechanism''.

\item  As expected from Schwinger's screening ideas \cite{schwinger}, \ The
physical $F_{\mu \nu }$- fields (those with commute with $Q)$ have vanishing
Maxwell charge and this would continue to be true in the presence of
additional spinor matter.

\item  The uniqueness of the renormalizable spin=1 part follows already from
the specification of the physical particle content \cite{D-Schr}; only the
coupling between s\TEXTsymbol{<}1 matter introduces the usual additional
parameters.
\end{itemize}

\textbf{Comments: }The results show that although the gauge point of view
which requires the Higgs-Kibble mechanism (``fattening of photons by eating
Goldstone Bosons'') is not incorrect, there is nothing physical-intrinsic
about it; it is a mnemotechnical device which allows to
differential-geometrically inclined physicist a rapid access to the
perturbative results. It has the disadvantage that the necessity of the
presence of additional physical degrees of freedom for reasons of
consistency within renormalized perturbation theory (the Higgs degree of
freedom) is not as convincing as in the present approach, in fact one
usually puts the Higgs fields into the Lagrangian from the beginning. The
present method leads to the same physical correlation functions but with a
slightly different conceptual ring. The ghosts are more clearly recognizable
as kinematical (via extension of $H_{Wig})$ auxiliary unphysical objects
whereas the dynamical presence of additional physical degrees of freedom
(the alias Higgs field, but without vacuum condensates) for matters of
perturbative consistency becomes more manifest and the observable particle
content receives greater emphasis. Classical differential geometric concepts
as the gauge idea are put into their proper place: they appear via Bohr's
correspondence principle on the classical side as a result of the uniqueness
of the implementation of perturbative renormalizability. Since gauge
theories play a very prominents role, this point of view is not without
interest. In fact it is close to the original viewpoint about massive
vectormesons by Sakurai. The idea of the BRS like cohomological extension
certainly takes care of those cases covered also by the gauge quantization
and the Higgs-Kibble mechanism, but it may have a larger range of
applicability to spin beyond one. The present method also suggests to
consider the conceptually simpler (validity of scattering theory) massive
case and approach the zero mass situation with its infrared problems as a
limiting case, i.e. the inverse of the Higgs approach. Since one knows that
the physical charge carrying fields in Maxwell-like theories have a
noncompact extension (spacelike cones with a semiinfinite string-like core),
the physical massive fields cannot converge without the necessity of a prior
modification. The attractive feature of such an idea is that such a
modification becomes related to the decoupling of the Higgs particle.

There is a special feature of abelian massive $s=1$ theories with additional
spinor matter which is absent in the nonabelian case. Namely in addition to
the massive theory constructed in tha analogous way with all matter fields
being renormalizable, there exists also ``massive QED'' for which the $\psi $%
-field cannot be simultaneously renormalizable (polynomially bounded
correlation function with a dominating degree independent on perturbative
order) and physical i.e. commuting with $Q.$ This massive QED has no Higgs
degree of freedom which is apparantly necessary in order to have both
properties.

A direct causal perturbative approach to s=1 massless theories was recently
formulated by Duetsch and Fredenhagen \cite{D-F}. The necessity to avoid the
(physically controversial) adiabatic limit requires the use of the full
nonlinear BRS structure and to confront a situation in which (unlike as in
the above case with bilinear $Q)$ the position of the physical cohomology
space keeps changing with the perturbative order. Lacking a fixed physical
reference space (e.g. an incoming scattering space) the physical space only
appears at the end as a representation space of a perturbative observable
*-algebra. This construction was carried out in QED, but there is little
doubt that with more work it also works for the nonabelian case.

We do of course not claim that the BRS-like cohomological construction for
the preservation of renormalizability in the face of higher spin advocated
in these notes is less mysterious then the quantization gauge principle. It
remains essentially unclear why and how the cohomological trick produces
local physical fields which at the end do not seem to be different from
those obtained with the standard causal perturbation method, except that the
latter cannot reconcile spin=1 with renormalizability. However it is a bit
closer to the spirit of LQP and perhaps less so to quantization and
differential geometry. It keeps the attention on the unsolved infrared
problems\footnote{%
From a physical point of view the estetical lure of differential geometry of
fibre bundles in gauge theories is a bit dangerous, because it takes one
away from the harder but physically more important infrared phenomena of the
LQP of s=1..} and it exposes the weird role of ghosts analogous to chemical
catalyzers: they are introduces into the original physical problem in order
to improve the $W$-powercounting and they are removed at the end without any
visible trace. The only difference to more standard renormalizable couplings
is the participation of $s=1$ vectormesons in the interaction vertex.

This situation cries out for a deeper understanding without ghosts. From the
more than 30 years struggle of physicist with this conceptual problem one
should conclude that if there exists a formulation without ghosts in
intermediate steps, than it cannot be anywhere near to the present
formulation. In fact the naturally ghostfree object is the S-matrix S which
in contradistiction to the above transition operator of the causal approach
S(g) is on-shell. If one could find an iteration scheme directly for S which
in intermediate steps avoids off-shell extrapolations than this would be
automatically ghostfree in every order. It would be a multivariable
dispersion theoretical approach based on unitarity and crossing symmetry.
The lowest order input consists of the on-shell tree diagrams (different
from the off-shell $W).$ Such an approach has only been carried out for
d=1+1 factorizing S-matrices where there exists a partial classification of
admissable S-matrices even without the use of perturbation theory: the
famous bootstrap program of factorizable models. Outside of such restrictive
situations a perturbative on-shell approach for $S$ does not yet exist. The
idea would be to use the perturbative ghostfree S-matrix in order to
construct polarization free generators of wedge algebras (PFG's). These are
operators which are similar to free fields in that their one time
application onto the vacuum is a one-particle vector without admixtures of
particle/antiparticle polarization clouds (see last section). In the
mentioned special case of factorizable models they are uniquely determined%
\footnote{%
In that case their Fourier transforms form a Zamolodchikov-Faddeev algebra 
\cite{Sch-Wie1}.} by the S-matrix via modular theory. Having generated the
wedge algebras from the S-matrix, one can than use modular ideas to define
and investigate a chiral conformal light ray theoy which is a canonical way
associated with the wedge algebra. Although many of these statements sound
futuristic, I think that this is the only way to avoid ghosts. One has to
bypass the use of a Wick-basis for the description of physical ghostfree
operators as linear combinations of composite fields. Such a basis is not
intrinsic and inevitably brings in the necessity of ghost field
contributions. The approach dealing with algebras is the only basis free
intrinsic approach to the problem. The difficulty is the conversion of these
rather abstract sounding ideas into concrete computational scheme. The
perturbative version of that only very incompletely understood scheme for
low-spin renormalizable models which did not need ghosts in the old
treatment should just reproduce the known renormalized results. Although our
main present motivation for going to such extremes was to have a ghostfree
renormalizable formalism for higher spin $s\geq 1,$ the interest in it would
by far exceed the present motivation. We will return to this issue of
generation of wedge algebras by modular methods in a more general context in
the last section.

\section{\protect\large Modular Origin of Geometric and Hidden Symmetries}

From the wedge localization in section 2 we have seen that the modular
objects associated to a standard (cyclic and separating vector $\Omega $)
pair ($\mathcal{A(O)},\Omega )$ has, under certain circumstances, a
geometrical significance, e.g. for the wedge in a massive
(Poincar\'{e}-invariant) theory, or the double cone in a massless
(conformally-invariant) theory. This suggests the question whether all
space-time symmetries (diffeomorphisms) can be viewed as having a modular
algebraic origin, i.e. if they can be thaught of as originating from the
relative positions of individual algebras in a net. \textit{This would
elevate spacetime from its role of merely indexing individual algebras in
the net, to a structure which is on the one hand more intimately related
with the physical aspects of LQP, and on the other hand emphasizes already
structural properties whose understanding seems to be a prerequisite for the
formulation of the elusive ``Quantum Gravity''.} It turns out that in chiral
conformal theories the Moebius group, together with the net on which it
acts, can be constructed from only two properly positioned algebras which
give rise to two ``halfsided modular inclusions'' (see below). In fact
mathematically the world of chiral conformal nets is equivalent with the
classification of all ``standard halfsided modular inclusions''. In this
conformal setting the Haag duality is automatic and there is no spontaneous
symmetry breaking. The analogue in the higher dimensional case is to assume
wedge duality (always achievable, as previously mentioned, by
maximalization) and to prove the equality of the modular group with the
Lorentz-boost without assuming (as Bisognano and Wichmann did) that the
algebras are generated by local fields. Presently this cannot be done
without making additional assumptions e. i. assumptions which cannot be
expressed in terms of modular positions only, but are suggested by
space-time geometry \cite{Sch-Wie2}. Amazingly one again succeeds to build
up the whole Poincar\'{e} group as well as the net from a small finite
number of algebras in appropriate modular positions (using modular
inclusions and modular intersections).

Since modular groups exist for each space time region one may ask about
their physical interpretation. Let us start with posing the opposite
question in a context where there are geometric candidates without obvious
modular origin. In chiral conformal theories one has a rich supply of
diffeomorphisms of the circle which have been around since the beginning of
the $70^{ies}.$ The way these mathematical structures were discovered by
physicist is somewhat bizarre and confusing. It is interesting to take a
brief look at history by permitting a short interlude, before presenting our
modular interpretation.

Apart from some early work of mathematicians (Gelfand, Fuchs) on
diffeomorphisms of $S^{1}$ and their associated Witt algebra (infinitesimal
diffeomorphisms without the central extension), the first observation by
physicist of this Witt algebra structure was made in the Veneziano dual
S-matrix model by Virasoro \cite{Vir}. Virasoro and followers realized that
the on-shell dual S-matrix model allowed for a nice off-shell presentation
in terms of a massless free field theory in d=1+1. Parallel to this, but
without any interrelation, there were detailed field theoretic
investigations of the representation of conformal generators in terms of the
energy momentum tensor $T$ and their action e.g. on the Thirring fields \cite
{Low-Schr} and the problem (formulated in Lowenstein's thesis and going back
to Greenberg) of classifying so-called ``Lie-fields'' \cite{Low}, the
predecessors of what in the rediscovered version 25 years later were called
W-fields. The next contribution came again from the dual model calculations
and consisted in the correct computation of the central term (for free
massless fermions) which was previously overlooked \cite{Ramond}. My own
contribution was the computation in 1973 of the general structure of the $T$-%
$T$ commutation relation in chiral conformal theories as a structural
consequence of translational covariance and causality which I presented
together with other results at the January 1974 V Brazilian Symposium in Rio
de janeiro\cite{Schroer 74}. Apart from not knowing the afore-mentioned free
fermion results, my motivation was quite different and consisted in the
search for nontrivial ``Lie field'' of which the energy momentum tensor was
the first illustration\footnote{%
The reason why many field theoretical results on low dimensional field
theories were only published in conference proceedings was sociological and
not scientific. Low-dimensional field theory for the benefit of higher
dimensional S-matrix models was considered of greater physical relevance
than its use as a theoretical laboratory for the test of general ideas on
interactions, a point of view which was later uphold by string theorist.}.
In the same year the conformal block decomposition was discovered (called
decomposition of local fields into nonlocal components) which solved the
Einstein ``causality paradox'' by noticing \cite{S-S} that local fields were
irreducible only with respect to a finite neighbourhood of the identiy but
not with respect to the center of the covering of $SL(2,R)\times SL(2,R).$
The illustration of this decomposition theory by nontrivial models (minimal
models) beyond exponential Bose fields had to wait for another 10 years \cite
{BPZ}. By that time the increased knowledge by physicist about infinite
dimensional Lie-algebras (affine algebras, diffeomorphism algebras) was
leaving its marks on low-dimensional QFT. This had besides many gains also
one disadvantage because the use of those infinite dimensional Lie-algebras
seperated these low-dimensional QFT sharply from higher dimensional standard
type of QFT to which such structures are not available. The modular point of
view which I will present in the sequel admits a higher dimensional analogue
and incorporates conformal and factorizing theories back into the framework
of general QFT.

Returning to the modular issue, let us look at a special subgroup whose
Lie-algebra is isomorphic to that of the Moebius group. Its action on the
circle is 
\begin{equation}
z\rightarrow \sqrt{\frac{a+bz^{2}}{c+dz^{2}}},\,\,\left( 
\begin{array}{ll}
a & b \\ 
c & d
\end{array}
\right) \in SU(1,1)
\end{equation}
where the cuts connecting both poles and zeros are chosen outside the unit
circle. In fact this defines a two-fold covering of the Moebius group. Given
an interval, its square root (inverse image of $z\rightarrow z^{2})$
consists of two disjoint intervals which are separately left invariant under
the above transformation group. The obvious conjecture is of course that (as
for the case of a single interval) the covering dilation subgroup is the
modular group of the pair ($\mathcal{A}(I_{1}\cup I_{2}),\Omega ).\,$But
this cannot be, because this action restricted to one interval is the same
as that of the dilation in the Moebius group but this, according to a
theorem by Takesaki \cite{Tak} this is not possible if the vacuum state
fulfills the Reeh-Schlieder property of being cyclic and separating for only
one interval. Since it never happens that two disjoint square root intervals
are contained in one interval of another such pair, there will be no
contradiction with the lack of the Reeh-Schlieder property for one interval.
A (quasifree) state on the Weyl algebra (which we take as an illustration of
a simple conformal model) which is invariant under the above covering
transformation \cite{Schr-Wies} is easily found in terms its two-point
function which belongs to the following scalar product: 
\begin{equation}
\left\langle f,g\right\rangle =\int \frac{f(x)g(y)}{\left[ \left( x-y\right)
\left( 1+xy\right) +i\varepsilon \right] ^{2}}(1+x^{2})(1+y^{2})dxdy
\end{equation}
where we used the linear presentation instead of the circular one ($SL(2,R)$
instead of $SU(1,1)$). This is to be compared with the standard inner
product belonging to the vacuum representation 
\begin{equation}
\left\langle f,g\right\rangle _{0}=\int \frac{f(x)g(y)}{(x-y+i0)^{2}}dxdy
\end{equation}
One easily checks that this inner product belongs to the same symplectic
form as the standard one namely 
\begin{equation}
\omega (f,g)=\func{Im}\left\langle f,g\right\rangle =\int fg^{\prime
}dx=\omega _{0}(f,g)=\func{Im}\left\langle f,g\right\rangle _{0}
\end{equation}
As for the standard case the criterium for a Fock representation is that the
inner product can be represented in terms of $\omega $ with the help of a
complex structure $I_{0},I_{0}^{2}=-1,$ with 
\begin{eqnarray}
\left\langle f,g\right\rangle _{0} &=&\omega (I_{0}f,g)=-\omega (f,I_{0}g) \\
\left( I_{0}f\right) (x) &\equiv &\int \frac{-1}{\left( x-y+i\varepsilon
\right) }f(y)dy  \nonumber
\end{eqnarray}
the analogous statement holds for $\left\langle f,g\right\rangle $ with $%
I_{0}$ replaced by $I$%
\begin{eqnarray}
I &=&\Gamma ^{-1}\circ I_{0}\circ \Gamma \\
\left( \Gamma f\right) (x) &\equiv &\int f(\frac{x}{2}+sign(x)\sqrt{1+\left( 
\frac{x}{2}\right) ^{2}})  \nonumber \\
\left\langle \Gamma f,\Gamma g\right\rangle _{0} &=&\left\langle
f,g\right\rangle  \nonumber
\end{eqnarray}
The changed inner product defines a changed quasifree state on the Weyl
algebra. The proof that the covering dilation 
\begin{eqnarray}
U(\lambda ) &=&\Gamma ^{-1}\circ V(\lambda )\circ \Gamma \\
(V(\lambda )f)(x) &=&f(\lambda x)  \nonumber
\end{eqnarray}
is indeed the modular group for the algebra of the disjoint intervals $\left[
-\infty ,-1\right] \cup \left[ 0,1\right] $ in this quasifree state, we only
have to check the appropriate KMS condition. From:

\begin{eqnarray}
\stackunder{\theta \uparrow 2\pi }{\lim } &<&U(\lambda )f,g>_{1}=\stackunder{%
\theta \uparrow 2\pi }{\lim }<V(\lambda )\circ \Gamma (f),\Gamma
(g)>=\left\langle \Gamma (g),\Gamma (f)\right\rangle = \\
&=&\left\langle g,f\right\rangle _{1}  \nonumber
\end{eqnarray}
one sees that the $U(\lambda )$ fulfils the KMS condition if both $f$ and $g$
are from one of the two intervals since $\Gamma $ transforms the space of $%
\left[ 0,1\right] $ localized functions into $\left[ -\infty ,-1\right] $
localized ones and vice versa.

This situation is very interesting, since although the chiral
diffeomorphisms allows no geometric generalization to diffeomorphisms in
higher dimensional LQP, the disconnected (and multiply connected) algebras
have modular groups which act in a non-pointlike manner inside these
disconnected local regions\footnote{%
Observable algebras in disconnected regions have also played a role as
indicators of the presence of charge sectors \cite{Rehren}\cite{Wasser}\cite
{schroer}.}. This is what we mean by ``hidden symmetries''. There is another
closely related aspect which strengthens the physical relevance of
disconnected regions. It was well-known for some time \cite{Haag} that such
situations break Haag duality i.e. 
\[
\mathcal{A}(\left( I_{1}\cup I_{2}\right) ^{\prime })\subset \mathcal{A}%
(I_{1}\cup I_{2})^{\prime } 
\]
if the net $\mathcal{A}$ has nontrivial superselection rules. For models
resulting from the maximal extension of the abelian current algebra the
mechanism which causes this obstruction against Haag duality has been
completely analyzed in \cite{schroer}. Very recently this has been
understood in complete generality (for rational theories i.e. those with a
finite number of sectors) in \cite{Kawa} by using very powerful methods of
subfactor theory. In the context of the above use of ``geometric states'',
one would conjecture that their lack of cyclicity leads to a Jones projector
which contains the information about the additional superselection sectors,
but this remains to be seen.

In the following we will look at two more illustrations of modular
constructions.

As a reference wedge we may take the wedge $W(l_{1},l_{2})$ spanned by the
light like vectors $l_{1,2}=e_{\pm }=(1,0,0,\pm 1),$ in which case we call
z,t the longitudinal and x,y the transversal coordinates (the light like
characterization of wedges is convenient for the following). This situation
suggests to decompose the Poincar\'{e} group generators into longitudinal,
transversal and mixed generators 
\begin{equation}
\,P_{\pm }=\frac{1}{\sqrt{2}}(P_{0}\pm
P_{z}),\,\,M_{0z};\,\,M_{12},\,\,P_{i};\,\,G_{i}^{(\pm )}\equiv \frac{1}{%
\sqrt{2}}(M_{i0}\pm M_{iz}),\,i=1,2
\end{equation}
The generators $G_{i}^{(\pm )}$ are precisely the ``translational'' pieces
of the euclidean stability groups $E^{(\pm )}(2)$ of the two light vectors $%
e_{\pm }$ which appeared in Wigner's representation theory for zero mass
particles. More recently these ``translations'' inside the homogenous
Lorentz group appeared in the structural analysis of ``Modular
Intersections'' of two wedges \cite{Borchers}\cite{Wies}. Apart from the
absence of the positive spectrum condition, its role is analogous to that of
the true translations $P_{\pm }$ with respect to halfsided ``Modular
Inclusions'' \cite{Wies}.

As one reads off from the commutation relations, $P_{i},G_{i}^{(+)},P_{\pm }$
have the interpretation of a central extension of a transversal ``Galilei
group''\footnote{%
This G's are only Galileian in the transverse sense; they tilt the wedge so
that one of the light like directions is maintained but the longitudinal
plane changes.} with the two ``translations'' $G_{i}^{(+)}$ representing the
Galilei generators, $P_{+}$ the central ``mass'' and $P_{-}$ the
``nonrelativistic Hamiltonian''. The longitudinal boost $M_{0z}$ scales the
Galilei generators $G_{i}^{(+)}$ and the ``mass'' $P_{+}.$ Geometrically the 
$G_{i}^{(+)}$ change the standard wedge (it tilts the logitudinal plane) and
the corresponding finite transformations generate a family of wedges whose
envelope is the halfspace $x_{-}\geq 0.$ The Galilei group together with the
boost $M_{0z}$ generate an 8-parametric subgroup $G^{(+)}(8)$ inside the
10-parametric Poincar\'{e} group\footnote{%
The Galileian group is usually introduced as a ``contraction'' of the
Poicar\'{e} group. But as the present discussion about the wedge (or rather
the following remarks about two wedges in a special modular intersection
position) shows , it also appears as a genuine subgroup of the Poincare
group. The latter fact seems to be less known.}: 
\begin{equation}
\,G^{(+)}(8):\,\,P_{\pm },\,\,M_{0z};\,\,M_{12},\,\,P_{i};\,\,G_{i}^{(+)}
\label{gen}
\end{equation}
The modular reflection $J$ transforms this group into an isomorphic $%
G^{(-)}(8).$

The Galileian group is usually introduced as a ``contraction'' of the
Poincar\'{e} group. But as the present discussion, the wedge (or rather as
in the following remarks, two wedges in a special modular intersection
position) shows , it also appears as a genuine subgroup of the Poincar\'{e}
group. The latter fact seems to be less known.

All observation have interesting generalizations to the conformal group in
massless theories in which case the associated natural space-time region is
the double cone.

This subgroup $G^{(+)}(8)$ is intimately related to the notion of modular
intersection see \cite{Borchers}\cite{Wies}. Let $l_{1},l_{2}$ and $l_{3}$
be 3 linear independent light like vectors and consider two wedges $%
W(l_{1},l_{2}),W(l_{1},l_{3})$ with $\Lambda _{12}$ and $\Lambda _{13}$ the
associated Lorentz boosts. As a result of this common $l_{1}$ the algebras $%
\mathcal{N}=\mathcal{A}(W(l_{1},l_{2})),\mathcal{M}=\mathcal{A}%
(W(l_{1},l_{3}))$ have a modular intersection with respect to the vector $%
\Omega .$ Then ($\mathcal{N\cap M})\subset \mathcal{M},\Omega )$ is a
so-called modular inclusion \cite{Wies}\cite{Araki}. Identifying $%
W(l_{1},l_{2})$ with the above standard wedge, we notice that the
longitudinal generators $P_{\pm },\,\,M_{0z}$ are related to the inclusion
of the standard wedge algebra into the full algebra $\mathcal{B}(H),$
whereas the Galilei generators $G_{i}^{(+)}$ are the ``translational'' part
of the stability group of the common light vector $l_{1}$ (i.e. of the
Wigner light-like little group).

To simplify the situation let us take d=1+2 with $\mathcal{G}(4),$ in which
case there is only one Galilei generator $G.\,$ In addition to the
``visible'' geometric subgroup of the Poincar\'{e} group, the modular theory
produces a ``hidden'' symmetry transformation $U_{\mathcal{N\cap M},\mathcal{%
M}}(a)$ which belongs to a region which is a intersection of two wedges:

\begin{equation}
U_{\mathcal{N\cap M},\mathcal{M}}(a):=\exp (\frac{ia}{2\pi }(\ln \Delta
_{N\cap M}-\ln \Delta _{M}))
\end{equation}
is a unitary group with positive generator. Moreover one has: 
\begin{equation}
U_{\mathcal{N\frown M},\mathcal{M}}(1-e^{-2\pi t})=\Delta _{\mathcal{M}%
}^{it}\Delta _{\mathcal{N\frown M}}^{-it}
\end{equation}
\begin{equation}
U_{\mathcal{N\cap M},\mathcal{M}}(e^{-2\pi t}a)=\Delta _{\mathcal{M}}^{it}U_{%
\mathcal{N\cap M},\mathcal{M}}(a)\Delta _{\mathcal{M}}^{-it}
\end{equation}
\begin{equation}
AdU_{\mathcal{N\cap M},\mathcal{M}}(-1)(\mathcal{M}\emph{)=}\mathcal{N}\cap 
\mathcal{M}
\end{equation}
and 
\begin{equation}
J_{\mathcal{M}}U_{\mathcal{N\frown M},\mathcal{M}}(a)J_{\mathcal{M}}=U_{%
\mathcal{N\frown M},\mathcal{M}}(-a).
\end{equation}
Similar results hold for $\mathcal{N}$\emph{\ }replacing $\mathcal{M}$\emph{%
. }Due to the intersection property we finally have the commutation relation 
\begin{equation}
\lbrack U_{\mathcal{N\frown M},\mathcal{M}}(a),U_{\mathcal{N\frown M},%
\mathcal{N}}(b)]=0
\end{equation}
which enables one to define the unitary group 
\begin{equation}
U_{\mathcal{N\frown M}}(a)=U_{\mathcal{N\frown M},\mathcal{M}}(-a)U_{%
\mathcal{N\frown M},\mathcal{N}}(a).
\end{equation}
This latter group can be rewritten as 
\begin{equation}
U_{\mathcal{N\frown M}}(1-e^{-2\pi t})=\Delta _{\mathcal{M}}^{it}\Delta _{%
\mathcal{N}}^{-it}
\end{equation}
and thereby recognized to be in our physical application the 1-parameter
Galilean subgroup $G$ (\ref{gen}) in the above remarks.

Now we notice that for $a<0$%
\begin{eqnarray}
AdU_{\mathcal{N\frown M},\mathcal{M}}(a)(\mathcal{M}\emph{)} &=&Ad\Delta _{%
\mathcal{M}}^{-i(\frac{1}{2\pi }\ln -a)}U_{\mathcal{N\frown M},\mathcal{M}%
}(-1)(\mathcal{M}) \\
&=&Ad\Delta _{\mathcal{M}}^{-i(\frac{1}{2\pi }\ln -a)}(\mathcal{N}\cap 
\mathcal{M})  \nonumber
\end{eqnarray}
Because $\Delta _{\mathcal{M}}^{it}$ acts geometrically as Lorentz boosts,
we have full knowledge of the geometrical action of $U_{\mathcal{N\frown M},%
\mathcal{M}}(a)$ on $\mathcal{M}$\emph{\ }for\emph{\ }$a<0.$ For $~a>0$ we
notice 
\begin{eqnarray}
AdU_{\mathcal{N\frown M},\mathcal{M}}(1)(\mathcal{M}\emph{)} &=&AdU_{%
\mathcal{N\frown M},\mathcal{M}}(2)(\mathcal{M}\cap \mathcal{N}\emph{)}=AdJ_{%
\mathcal{M}}J_{\mathcal{N\frown M}}(\mathcal{M}\cap \mathcal{N}\emph{)} \\
&=&AdJ_{\mathcal{M}}(\mathcal{M}^{\prime }\cup \mathcal{N}^{\prime }) 
\nonumber
\end{eqnarray}
and again, due to the geometrical action of $J_{M}$ we have a geometrical
action on $\mathcal{M}$ for $a>0.$%
\begin{equation}
AdU_{\mathcal{N\cap M},\mathcal{M}}(a)(\mathcal{M}\emph{)=}Ad\Delta _{%
\mathcal{M}}^{-i(\frac{1}{2\pi }\ln a)}J_{\mathcal{M}}(\mathcal{M}^{\prime
}\cup \mathcal{N}^{\prime })
\end{equation}
From these observations and with $U_{\mathcal{N\cap M},\mathcal{M}%
}(1-e^{-2\pi t})=\Delta _{\mathcal{M}}^{it}\Delta _{\mathcal{M\cap N}}^{-it}$
we get for $t<0:$%
\begin{equation}
Ad\Delta _{\mathcal{N\cap M}}^{it}(\mathcal{M})=Ad\Delta _{\mathcal{M}}^{(-%
\frac{i}{2\pi }\ln (e^{-2\pi t}-1))}J_{\mathcal{M}}(\mathcal{M}^{\prime
}\cup \mathcal{N}^{\prime })
\end{equation}
and in case of $t>0:$%
\begin{equation}
Ad\Delta _{\mathcal{N\cap M}}^{it}(\mathcal{M})=Ad\Delta _{\mathcal{M}}^{(-%
\frac{i}{2\pi }\ln (1-e^{-2\pi t}))}(\mathcal{N}\cap \mathcal{M}).
\end{equation}
Similar results hold for $\mathcal{N}$\emph{\ }~replacing $\mathcal{M}$\emph{%
\ . }With the same methods we get:\smallskip 
\begin{eqnarray}
Ad\Delta _{\mathcal{N\cap M}}^{it}\Delta _{\mathcal{N}}^{is}(\mathcal{M})
&=&Ad\Delta _{\mathcal{N\cap M}}^{it}\Delta _{\mathcal{N}}^{is}\Delta _{%
\mathcal{M}}^{-is}(\mathcal{M}) \\
&=&Ad\Delta _{\mathcal{N\cap M}}^{it}U_{\mathcal{M\cap N}}(e^{-2\pi s}-1)(%
\mathcal{M})  \nonumber
\end{eqnarray}
where $U_{\mathcal{N\cap M}}$ is the 1-parameter Lorentz subgroup (the
Galilei subgroup $G$ in (\ref{gen}) associated with the modular
intersection. This gives: 
\begin{eqnarray}
Ad\Delta _{\mathcal{N\cap M}}^{it}\Delta _{N}^{is}(\mathcal{M}) &=&AdU_{%
\mathcal{M\cap N}}(e^{-2\pi t}(e^{-2\pi s}-1))\Delta _{\mathcal{N\cap M}%
}^{it}(\mathcal{M}) \\
&=&AdU_{\mathcal{M\cap N}}(e^{-2\pi t}(e^{-2\pi s}-1))\Delta _{\mathcal{M}%
}^{-\frac{1}{2\pi }\ln (1-e^{-2\pi t})}(\mathcal{M\cap N}),  \nonumber
\end{eqnarray}
if $t>0$ and similar for $t<0$.Therefore we get a geometrical action of $%
\Delta _{\mathcal{N\cap M}}^{it}$ on $Ad\Delta _{\mathcal{N}}^{is}(\mathcal{M%
}).$

A look at the proof shows that the essential ingredients are the special
commutation relations. Due to 
\begin{equation}
\Delta _{\mathcal{M\cap N}}^{it}=\Delta _{\mathcal{M}}^{it}U_{\mathcal{N\cap
M},\mathcal{M}}(1-e^{-2\pi t})=\Delta _{\mathcal{M}}^{it}J_{\mathcal{M}}U_{%
\mathcal{N\cap M},\mathcal{M}}(e^{-2\pi t}-1)J_{\mathcal{M}}
\end{equation}
and the well established geometrical action of $\Delta _{\mathcal{M}}^{it}$
and $J_{\mathcal{M}},$ it is enough to consider the action of $U_{\mathcal{%
N\cap M},\mathcal{M}}$ or similarly $U_{\mathcal{N\cap M},\mathcal{N}}.$ For
these groups we easily get 
\begin{equation}
AdU_{\mathcal{N\cap M},\mathcal{M}}(a)\Delta _{\mathcal{N}}^{is}\Delta _{%
\mathcal{M}}^{-it}(\mathcal{N})=Ad\Delta _{\mathcal{N}}^{is}\Delta _{%
\mathcal{M}}^{it}U_{\mathcal{N\cap M},\mathcal{M}}(e^{-2\pi (s+t)}a)(%
\mathcal{N)}
\end{equation}
and due to the above remarks the geometrical action of $\Delta _{\mathcal{%
N\cap M}}^{it}$ on the algebras of the type $Ad\Delta _{\mathcal{N}%
}^{is}\Delta _{\mathcal{M}}^{-it}(\mathcal{M})$.

Now, the lightlike translations $U_{transl_{1}\,}(a)$ in $l_{1}$ direction
fulfill the positive spectrum condition and map $\mathcal{N}\cap \mathcal{M}$
into itself for $a>0.$ Therefore we have the Borchers commutator relations
with $\Delta _{\mathcal{M\cap N}}^{it}$ and get 
\begin{equation}
Ad\Delta _{\mathcal{N\frown M}}^{it}U_{transl_{1}}(a)(\mathcal{M}%
)=AdU_{transl_{1}}(e^{-2\pi t}a)\Delta _{\mathcal{N\frown M}}^{it}(\mathcal{M%
})
\end{equation}
The additivity of the net tells us that taking unions of the algebra
corresponds to the causal unions of localization regions. The assumed
duality allows us to pass to causal complements and thereby to intersections
of the underlying localization regions. Therefore the algebraic properties
above transfer to unions, causal complements and intersections of regions.
We finally get \cite{Schr-Wies}:

\begin{theorem}
Let $\mathcal{R}$ be the set of regions in $\mathbf{R}^{1,2}$ containing the
wedges $W[l_{1},l_{2}],$

$W[l_{1},l_{3}]$ and which is closed under:

a) Lorentz boosting with $\Lambda _{12}(t),\Lambda _{13}(s),$

b) intersection

c) (causal) union

d) translation in $l_{1}$ direction

e) causal complement

Then $\Delta _{W[l_{1},l_{2}]\cap W[l_{1},l_{3}]}^{it}$ maps sets in $%
\mathcal{R}$ onto sets in $\mathcal{R}$ in a well computable way and extends
the subgroup (\ref{gen}) by a ``hidden symmetry''.
\end{theorem}

Similarly we can look at a (1+3)-dim. quantum field theory. Then we get the
same results as above for the modular theory to the region $%
W[l_{1},l_{2}]\cap W[l_{1},l_{3}]\cap W[l_{1},l_{4}],$ where $l_{i}$ are 4
linear independent lightlike vectors in $\mathbf{R}^{1,3}.$ Moreover in this
case the set $\mathcal{R}$ contain $W[l_{1},l_{2}],W[l_{1},l_{3}]$ and $%
W[l_{1},l_{4}]$ and is closed under boosting with $\Lambda _{12}(t),\Lambda
_{13}(s),\Lambda _{14}(r).$

The arguments are based on the Borchers commutation relation and modular
intersection theory and apply also if we replace modular intersection by
modular inclusion. One recovers in this way easily the results of Borchers
and Yngvason, \cite{Bo/Yng} who found an illustration of hidden symmetries
in thermal chiral conformal QFT ( Note that in thermal situations we have no
simple geometrical interpretation for the commutants as the algebra to
causal complements. Therefore in these cases we have to drop e) in the above
theorem.).

The final upshot of this section is to show that there might be a well
defined meaning of a geometrical action of modular groups by restricting on
certain subsystems.

For conformal LQP in any dimension, one obtains a generalization of the
previous situation. In particular the modular group with respect to the
vacuum of the double cone algebra is geometric\cite{Haag}. Consider now a
double cone algebra $\mathcal{A}(\mathcal{O})$ generated by a free massless
field (for s=0 take the infrared convergent derivative). Then according to
the previous remark, the modular objects of ($\mathcal{A}(\mathcal{O}%
),\Omega )_{m=0}$ are well-known . In particular the modular group is a one
parametric subgroup of the proper conformal group. The massive double cone
algebra together with the (wrong) massless vacuum has the same modular group 
$\sigma _{t}$ however its action on smaller massive subalgebras inside the
original one is not describable in terms of the previous subgroup. In fact
the geometrical aspect of the action is wrecked by the breakdown of Huygens
principle, which leads to a nonlocal reshuffling inside $\mathcal{O}$ but
still is local in the sense of keeping the inside and its causal complement
apart. This mechanism can be shown to lead to a pseudo-differential operator
for the infinitesimal generator of $\sigma _{t}$ whose's highest term still
agrees with conformal zero mass differential operator. We are however
interested in the modular group of ($\mathcal{A}(\mathcal{O}),\Omega )_{m}$
with the massive vacuum which is different from the that of the wrong vacuum
by a Connes cocycle. We believe that this modular cocycle will not wreck the
pseudo-differential nature and that as a consequence the geometric nature of
the conformal situation will still be asymptotically true near the horizon
of the double cone, however we were presently not able to show this. This
modular aspect of the horizon could be linked with what people think should
be the quantum version of the Bekenstein-Hawking classical entropy
considerations, in particular the ideas about ``holographic properties''. To
be more precise, we expect that even for double cones in Minkowski space
(i.e. without a classical Killing vector as for black holes) there will be a
finite relative quantum entropy as long as one allows for a ``collar''
between the double cone and its spacelike complement and that with vanishing
size of this collar these entropies will diverge in such a way that ratios
(e.g. for differently sized double cones) will stay finite and be determined
by the conformal limits. In this way one could hope to prove that e.g. the
speculations about entropy, holography and the occurrence of the central
terms in the energy momentum commutation relations are nonperturbative
generic properties of ordinary LQP \cite{Sch Lec}. For the thermal aspects
this is of course well known..

The modular group structure also promise to clarify some points concerning
the physics of the Wightman domain properties \cite{Sch AOP}. In fact these
groups act linearly on the ''field space'' i. e. the space generated by
applying a local field on the vacuum. Therefore this space, which is highly
reducible under the Poincar\'{e} group, may according to a conjecture of
Fredenhagen (based on the results in \cite{Fred-Joe}) in fact carry an
irreducible representation of the union of all modular groups (an infinite
dimensional group $\mathcal{G}_{mod}$ which contains in particular all local
spacetime symmetries). The equivalence of fields with carriers of
irreducible representations of an universal $\mathcal{G}_{mod}$ would add a
significant conceptual element to LQP and give the notion of quantum fields
a deep role which goes much beyond that of being simply generators of local
algebras. Our arguments suggest that in chiral conformal QFT $\mathcal{G}%
_{mod}$ includes all local diffeomorphism.

A related group theoretical approach to LQP which uses both modular groups
and modular involutions in order to formulate a new selection principle for
states (''The Condition of Geometric Modular Action'') was proposed in \cite
{Buch et al}. In addition to the modular groups which leave the defining
local algebras invariant, these authors obtain a discrete group (from the
conjugations) which transform the (spacetime) index set. All these true QFT
properties remain invisible in any quantization approach. Combining modular
theory with scattering theory, the actual $J$ together with the incoming $%
J^{in}$ can be used to obtain a new framework for nonperturbative
interactions \cite{Sch AOP}. This last topic will be presented in the
following section; more details can be found in a separate paper together
with H.-W. Wiesbrock \cite{Sch-Wie1}.

\section{\protect\large Constructive Modular Approach to Interactions}

The starting observation for relating the modular structure of LQP nets to
interactions is that the latter is solely contained in those anti-unitary
reflections of the full Poincar\'{e} group which contain the time reversal.
The continuous part (as well as those reflections which do not involve time)
is, thanks to the fact that scattering (Haag-Ruelle, LSZ) theory is a
consequence of LQP, the same for the free incoming particles as for the
interacting net \cite{schroer}: 
\begin{eqnarray}
U(\Lambda ,a) &=&U(\Lambda ,a)^{in} \\
J &=&S_{sc}J^{in}  \nonumber \\
S_{T} &=&J\Delta ^{\frac{1}{2}}  \nonumber \\
S_{T}A\Omega &=&A^{\ast }\Omega
\end{eqnarray}
Here $S$ is the scattering matrix. The subscript $T$ is used in order to
distinguish the Tomita, operator from the scattering matrix and the $J$ is
the Tomita reflection for interacting wedge algebras whereas $J^{in}$ refers
to the algebra generated by the incoming free field. The standard point of
view, where the interaction is introduced in terms of a pair of Hamiltonians
(Lagrangians) $H,H_{0},$ accounts for the interaction in another (more
perturbative) way which uses different states. It is well-known that this
standard perturbative approach cannot be directly formulated in infinite
space because translational invariance together with invariance of the
vacuum is in contradiction with the existence of another hamiltonian $H$
once a bilinear $H_{0}$ has been specified (Haag's theorem). In perturbation
theory this is not a series obstacle; it is formally taken care of by
leaving out the pure vacuum Feynman graphs or more carefully by using the
Feynman-Gell-Mann formula in a quantization box and taking the thermodynamic
limit. The modular approach does not have this problem.

The most promising candidates for a modular construction are obviously
massive theories with a known S-Matrix i.e. models which permit a bootstrap
construction of S on its own, without using the off-shell fields or local
operators. For such S-matrix integrable models, there already exists a
constructive formfactor program which goes back to Karowski and Weisz and
has been significantly extended by Smirnov \cite{Ka-Wei}\cite{Smir}. It uses
suggestive prescriptions and assumptions within the dispersion theoretical
LSZ framework.

Since the bulk of the LSZ formalism is a consequence of the more basic
algebraic QFT, it is reasonable to ask if our modular localization framework
is capable to shed additional light on this program in particular whether it
can be understood as a special (analytically simple) case of a more general
nonperturbative construction without the restriction to d=1+1 factorizing
theories \cite{Sch AOP}. The crucial vehicle which carries the off-shell
modular and thermal properties of wedge regions to on-shell crossing
properties of formfactors are very subtle polarization-free wedge generators
(PFG) which we will now explain.

Let us start with a very simple-minded generalization of free fields in
d=1+1. For the latter we use the notation: 
\begin{eqnarray}
A(x) &=&\frac{1}{\sqrt{2\pi }}\int (e^{-ipx}a(p)+h.a.)\frac{dp}{2\omega } \\
&=&\frac{1}{\sqrt{2\pi }}\int (e^{-im\rho sh(\chi -\theta )}a(\theta
)+h.a.)d\theta ,\,\,x^{2}<0  \nonumber \\
&=&\frac{1}{\sqrt{2\pi }}\int_{\mathsf{C}}e^{-im\rho sh(\chi -\theta
)}a(\theta )d\theta ,\,\,\,\mathsf{C}=\Bbb{R\cup }\left\{ -i\pi +\Bbb{R}%
\right\} \mathsf{\,\,\,}  \nonumber
\end{eqnarray}
where in the second line we have introduced the x- and momentum- space
rapidities and specialized to the case of spacelike x, and in the third line
we used the analytic properties of the exponential factors in order to
arrive at a compact and (as it will turn out) useful contour representation.
Note that the analytic continuation refers to the c-number function, whereas
the formula $a(\theta -i\pi )\equiv a^{\ast }(\theta )$ is a definition and
has nothing to do with analytic continuations of operators\footnote{%
Operators in QFT never possess analytic properties in x- or p-space. The
notation and terminology in conformal field theory is a bit confusing on
this point, because although it is used for operators it really should refer
to vector states and expectation values in certain representations of the
abstract operators. The use of modular methods require more conceptual
clarity than standard methods.}.

With this notational matter out of the way, we now write down our Ansatz 
\begin{eqnarray}
F(x) &=&\frac{1}{\sqrt{2\pi }}\int_{\mathsf{C}}e^{-im\rho sh(\chi -\theta
)}Z(\theta )d\theta \\
Z(\theta )\Omega &=&0,\,\,\,Z(\theta _{1})Z(\theta _{2})=S_{Z,Z}(\theta
_{1}-\theta _{2})Z(\theta _{2})Z(\theta _{1})  \label{Ansatz} \\
Z(\theta _{1})Z^{\ast }(\theta _{2}) &=&\delta (\theta _{1}-\theta
_{2})+S_{Z,Z^{\ast }}(\theta _{1}-\theta _{2})Z^{\ast }(\theta _{2})Z(\theta
_{1})  \nonumber
\end{eqnarray}
For the moment the $S^{^{\prime }}s$ are simply Lorentz-covariant (only
rapidity differences appear) functions which for algebraic consistency
fulfil unitarity $\overline{S(\theta )}=S(-\theta ).$ We assume (for
simplicity) that the state space contains only one type of particle.

\bigskip A field operator F(x) is called ``one-particle \textbf{p}%
olarization \textbf{f}ree'' (PF) if F(x)$\Omega $ and F$^{\ast }$(x)$\Omega $
have only one-particle components (for any one of the irreducible particle
spaces in the theory)

Obviously the above $F(x)$ with $Z(\theta )\Omega =0$ (but yet without the
algebraic relations which specialize the interactions to the relativistic
counterpart of quantum mechanical pair interactions) is the most general PF
in d=1+1. The PF property is an on-shell concept , but note that nothing is
required about the nature of state vectors which are created by several
PF's. As a result of an old structural theorem of QFT, a PF is pointlike
local, if and only if it is a free field \cite{ST-Wi}, i.e. if and only if
the Fourier-components $Z^{\#}(\theta )$ fulfil the free field commutation
relation which coalesce with those of the above Ansatz for $%
S_{Z,Z}=1=S_{Z,Z^{\ast }}.$ Although interacting PF's are necessarily
nonlocal, it is an interesting question how nonlocal they must be in order
not to fall under the reign of the structural theorem. It turns out that
they can be localized in wedges but any sharper localization requirement
reduces them to free fields. In the more \textit{special context of the
above Ansatz} we find \cite{Sch-Wie1}

\begin{proposition}
The requirement of wedge localization of a PF operator $F(f)=\int
F(x)f(x)d^{2}x,\,suppf\in W$ is equivalent to the Zamolodchikov-Faddeev
structure of the Z-algebra. The corresponding F's cannot be localized in
smaller regions i.e. the localization of F(f) with suppf$\in \mathcal{O}%
\subset W$ is not in $\mathcal{O}$ but still uses all of $W.$
\end{proposition}

Before doing the necessary calculation, let us put on record two more
definitions of a general kind which are suggested by the proposition.

\begin{definition}
We call PF's which generate the wedge algebra\footnote{%
In this letter we do not discuss the necessity to distinguish between
localized von Neumann algebras $\mathcal{A}(\mathcal{O})$ of bounded
operators and polynomial algebras $\mathcal{P}(\mathcal{O})$ of affiliated
unbounded operators as those formed from products of $F(f)$'s and their
precise relation.} 
\[
\mathcal{A}(W)=alg\left\{ F(\hat{f}),\forall f\,\,supp\hat{f}\in W\right\}
\]

PFG or one-particle \textbf{p}olarization \textbf{f}ree wedge \textbf{g}%
enerators \cite{Sch-Wie1}.
\end{definition}

We omitted the w for wedge in our short hand notation because on the one
hand wedges are the ``smallest'' regions in Minkowski space which do not
have the full space as the causal closure and possess PF's. In view of the
fact that we work more frequently in momentum space and its
rapidity-parametrized mass-shell restriction (often referred to as
one-particle wave functions), we reserve the simpler notation f without hat
to the Fourier transforms.

\begin{definition}
We call the improvement of localization obtained by intersecting $\mathcal{A}%
(W)^{\prime }s$ for different wedges an improvement of \ ``quantum
localization'' \cite{Sch-Wie1}, whereas the standard localization in $suppf$ with
the use of smeared out pointlike local fields $A(f)$ is referred to as
classical (albeit in a \textbf{quantum} field theory).
\end{definition}

We now prove the proposition by employing the so called KMS condition for
localized algebras. This property originally arose in thermal systems in
cases where the thermodynamical limit for the infinitely extended system
cannot be described in terms of a Gibbs formula (volume divergencies), but
it later turned out to be generally valid for all systems which result von
Neumann algebras $\mathcal{A}$ in a cyclic and separating state vector $%
\Omega :$%
\begin{equation}
\left( \Omega ,A\sigma _{t}(B)\Omega \right) =\left( \Omega ,\sigma
_{t+i}(B)A\Omega \right)
\end{equation}
where $\sigma _{t}(B)\equiv Ad\Delta ^{it}(B)$ is the action of the modular
group. Local algebras in QFT are known to have this commutation property
with respect to the vacuum state at least as long as the localization region
has a nontrivial causal complement, but they generally do not admit a
natural thermodynamic limit description in terms of a sequence of increasing
quantization boxes. For the wedge regions at hand, the localized field
algebras are known to have the wedge affiliated Lorentz boost as their KMS
automorphism group $\sigma _{t}$.

\begin{proof}
Consider first the KMS property of the two-point function 
\begin{equation}
\left\langle F(f_{1})F(f_{2})\right\rangle =\left\langle F(f_{2}^{2\pi
i})F(f_{1})\right\rangle =\left\langle F(f_{2}^{\pi i})F(f_{1}^{-i\pi
})\right\rangle
\end{equation}
Rewritten in terms of the f's we have 
\begin{equation}
\int f_{1}(\theta )\bar{f}_{2}(\theta )d\theta =\int f_{2}(\theta -i\pi )%
\bar{f}_{1}(\theta +i\pi )d\theta
\end{equation}
which is an identity in view of the fact that the wedge support properties
for the test functions f together with their reality condition imply $%
f(\theta -i\pi )=\bar{f}(\theta ).$

The 4-point function $\left\langle 1,2,3,4\right\rangle $ consists of 3
contributions, one from an intermediate vacuum state vector associated with
the contraction scheme $\left\langle 12\right\rangle \left\langle
34\right\rangle ,$ another one from the direct intermediate two-particle
contribution $\left\langle 14\right\rangle $ $\left\langle 23\right\rangle $%
and the third one from its exchanged (crossed) version $\left\langle
13\right\rangle \left\langle 24\right\rangle .$ The latter is the only one
which carries the interaction in form of the $S$-coefficients. In the would
be KMS relation 
\begin{eqnarray}
\left\langle F(f_{1})F(f_{2})F(f_{3})F(f_{4})\right\rangle &=&\left\langle
F(f_{4}^{-2\pi i})F(f_{1})F(f_{2})F(f_{3})\right\rangle \\
f^{z}(\theta ) &:&=a.c.f\mid _{\theta \rightarrow \theta +z}  \nonumber
\end{eqnarray}
the vacuum terms and the direct terms interchange their role on both sides
of the equation and cancel out, whereas the crossed terms are related by
analytic continuation. The required equality for the crossed term brings in
the S-matrix via the relations (\ref{Ansatz}) and yields 
\begin{eqnarray}
&&\int \int d\theta d\theta ^{\prime }S(\theta -\theta ^{\prime
})f_{2}(\theta )\bar{f}_{4}(\theta )f_{1}(\theta ^{\prime })\bar{f}%
_{3}(\theta ^{\prime }) \\
&=&\int \int d\theta d\theta ^{\prime }S(\theta -\theta ^{\prime
})f_{1}(\theta )\bar{f}_{3}(\theta )f_{4}(\theta ^{\prime }-2\pi i)\bar{f}%
_{2}(\theta ^{\prime })  \nonumber
\end{eqnarray}
Again using the above boundary relation for the wave functions we rewrite
the last product in the second line as $\bar{f}_{4}(\theta ^{\prime }-i\pi
)f_{2}(\theta ^{\prime }-i\pi )$ and performing a contour shift $\theta
^{\prime }\rightarrow \theta ^{\prime }+i\pi ,$ renaming $\theta
\leftrightarrow \theta ^{\prime }$ and finally using the denseness of the
wave functions in the Hilbert space, we obtain the crossing relation for $S$ 
\begin{equation}
S(\theta )=S(-\theta +i\pi )
\end{equation}
Note that we already omitted the subscripts on S, since the identity $%
S_{Z,Z^{\ast }}=S_{Z,Z}\equiv S$ follows from the two different ways of
calculating the crossed term, once by interchanging the two creation
operators in $Z^{\ast }(\theta _{3})Z^{\ast }(\theta _{4})$ and then
performing the direct contraction and another way by interchanging $Z(\theta
_{2})Z^{\ast }(\theta _{3})$ and then being left with the vacuum
contraction. Let us look at one more KMS relation for the six-point
functions of the would be PFG's. 
\begin{equation}
\left\langle F(f_{1})....F(f_{6})\right\rangle =\left\langle F(f_{6}^{2\pi
i})F(f_{1})...F(f_{5})\right\rangle
\end{equation}
This time one has many more pairings In fact ordering with respect to pair
contraction times 4-point functions one may again group the various terms in
those for which the pairing contraction is between adjacent $Z^{\prime }s$
and those where this only can be achieved by exchanges. The first group
satisfies the KMS condition because of the previous verification for the 2-
and 4- point functions. For the crossed contributions the wave functions say 
$f_{i}$ and $\bar{f}_{k}.$ Those terms only compensate by shifting upper 
\textsf{C-}contours into lower ones and vice versa. If S would contain poles
in the physical sheet, then there are additional contributions and the KMS
property only holds if these poles occur in symmetric pairs i.e. in a
crossing symmetric fashion.
\end{proof}

We will not pursue the fusion structure for the Z's resulting from poles
beyond noting that the particle spectrum already shows up in the fusion of
the wedge localized $Z(f)^{\prime }s.$ One of course expects agreement of
the fusion structure of our PFG's with the formal Zamolodchikov conjecture%
\footnote{%
In fact it is only through the PFG's F(x) that the Z-F algebra and the
fusion rules for the Z's receive a space-time interpretation. The close
relation to a kind of relativistic QM only happens on the level of wedge
localization; the algebras resulting from intersections of wedge algebras
loose this quantum mechanical aspect and show the full virtual particle
creation/annihilation polarization structure.}, however a detailed
discussion of fusion would go beyond the aim of this letter and will be the
subject of a separate paper. It should be stressed that the simple quantum
mechanical picture of fusion in terms of bound states only holds for the
above model with pair interactions and not for more realistic models with
real (on-shell) particle creation. All models whether they are real particle
conserving or not (except free fields) have a rich virtual particle
structure (as will shown later), i.e. the particle content of operators $A$
with compact localization e.g. \ $A\in \mathcal{A}(\mathcal{O})$ complies
with the ``folklore'' that all particle matrix elements 
\begin{equation}
^{out}\left\langle p_{1},...,p_{k}\left| A\right|
q_{1},...,q_{l}\right\rangle ^{in}\neq 0
\end{equation}
as long as they are not forced to vanish by superselection rules.

Although we have explained the basic concepts in the case of diagonal
S-coefficients in the Z-algebra, one realizes immediately that one can
generalize the formalism to \textit{matrix-valued} ``pair interactions'' S.
The operator formalism (the associativity) then leads to the Yang-Baxter
conditions and the crossing relations are again equivalent to the KMS
property for the wedge generators $F(f)$.

The relation of the above observation with \textbf{l}ocal \textbf{q}uantum 
\textbf{p}hysics (LQP) becomes more tight, if one remembers that the Lorentz
boost, which featured in the above KMS condition, also appears together with
the TCP operator in the Tomita modular theory for the pair ($\mathcal{A}%
(W),\Omega $): 
\begin{equation}
S_{T}A\Omega =A^{\ast }\Omega ,\,\,\,A\in \mathcal{A}(W)
\end{equation}
which defines the antilinear, unbounded, closable, involutive (on its
domain) Tomita operator $S_{T}.$ Its polar decomposition 
\begin{equation}
S_{T}=J\Delta ^{\frac{1}{2}}
\end{equation}
defines a positive unbounded $\Delta ^{\frac{1}{2}}$ and an antiunitary
involutive $J$ and the nontrivial part of Tomita's theorem (with
improvements by Takesaki) is that the unitaty $\Delta ^{it}$ defines an
automorphism of the algebra i.e. $\sigma _{t}(\mathcal{A})\equiv \Delta ^{it}%
\mathcal{A}\Delta ^{-it}=\mathcal{A}$ and the $J$ maps into antiunitarily
into its commutant $j(\mathcal{A})\equiv J\mathcal{A}J=\mathcal{A}^{\prime
}. $ The wedge situation is a special illustration for the Tomita theory. In
that case both operators are well-known; the modular group is the
one-parametric wedge affiliated Lorentz boost group $\Delta ^{it}=U(\Lambda
(-2\pi t),$ and the $J$ in d=1+1 LQP's is the fundamental TCP-operator (in
higher dimensions it is only different by a $\pi $-rotation around the
spatial wedge axis). The prerequisite for the general Tomita situation is
that the vector in the pair (algebra, vector) is cyclic and separating (no
annihilation operators in the von Neumann algebra resp. cyclicity of \ its
commutant relative to the reference vector). In LQP these properties are
guarantied for localization regions $\mathcal{O}$ with nontrivial causal
complement $\mathcal{O}^{\prime }$ thanks to the Reeh-Schlieder theorem.
Returning to our wedge situation we conclude from the Bisognano-Wichmann
result that the commutant of $\mathcal{A}(W)$ is geometric i.e. fulfils Haag
duality $\mathcal{A}(W)^{\prime }=\mathcal{A}(W^{\prime }),$ a fact which
can be shown to be modified by Klein factors in $J$ in case of deviation
from Bose statistics.

There is one more structural element following from ``quantum localization''
beyond wedge localization.

\begin{proposition}
Operators localized in double cones $A\in A(\mathcal{O})$ obey a recursion
relation in their expansion coefficients in terms of PFG operators 
\begin{eqnarray*}
&&A=\sum \frac{1}{n!}\int_{\mathsf{C}}...\int_{\mathsf{C}}a_{n}(\theta
_{1},...\theta _{n}):Z(\theta _{1})...Z(\theta _{n}):d\theta _{1}...d\theta
_{n} \\
&=&\sum \frac{1}{n!}\int ...\int \hat{a}%
_{n}(x_{1},...x_{n}):F(x_{1})...F(x_{n}):d^{2}x_{1}...d^{2}x_{n},\,\,supp%
\hat{a}\in W^{\otimes n} \\
&&ilim_{\theta \rightarrow \theta _{1}}(\theta -\theta _{1})a_{n+1}(\theta
,\theta _{1},...,\theta _{n})=(1-\prod_{i=2}^{n}S(\theta _{1}-\theta
_{i}))a_{n-1}(\theta _{2},..,\theta _{n})
\end{eqnarray*}
\end{proposition}

\begin{remark}
In order to compare (see below) with Smirnov's \cite{Smir} axioms we wrote the
recursion in rapidity space instead of in x-space light-ray restriction
which would be more physical and natural to our modular approach. The series
extends typically to infinity. Only for special operators (e.g. bilinears as
the energy momentum tensor) in special models with rapidity independent
S-matrices (e.g. Ising, Federbush) for which the bracket involving the
product of two-particle S-matrices vanishes, the series restricts to a
polynomial expression in Z. Therefore apart from these special cases, an
operator $A\in \mathcal{A}(\mathcal{O})$ with $a_{1}\neq 0$ applied to the
vacuum creates a one-particle component which an admixture of an infinite
cloud of additional particles (particle-antiparticle polarization cloud).
The above recursion together with Payley-Wiener type bounds for the increase
of the $a_{n}^{\prime }s$ in imaginary $\theta $-directions (depending on
the shape and size of $\mathcal{O)}.$
\end{remark}

The prove follows rather straightforwardly from the quantum localization
idea 
\begin{equation}
\mathcal{A}(\mathcal{O})=\left[ U(a)\mathcal{A}(W)U^{-1}(a)\right] ^{\prime
}\cap \mathcal{A}(W)
\end{equation}
i.e. we are considering the relative commutant inside the wedge algebra.
Using the PFG's $F(f),$ the $A\in \mathcal{A}(\mathcal{O})$ are
characterized by \cite{Sch-Wie1} 
\begin{equation}
\left[ A,F(\hat{f}_{a})\right] =0,\,\,\forall \hat{f}\in W
\end{equation}
where $\hat{f}_{a}(x)=\hat{f}(x-a),\,a\in W.$ One immediately realizes that
the contribution of the commutator to the $n^{th}$ power in $F$ yields a
relation between the $a_{n-1}$ and $a_{n+1}$ (from the creation/annihilation
part of $\ F(\hat{f}_{a})).$ The details of this relation are easier, if one
passes to the light-ray restriction which in the present approach turns out
to be a very nontrivial result of modular theory \cite{Sch-Wie1}\cite
{Sch-Wie2}\cite{GLRV}.

\begin{proposition}
The relative commutant for light-like translations with $a_{+}=(1,1)$
defines a ``satellite'' chiral conformal field theory via the (half) net on
the (upper) +light ray 
\begin{equation}
\mathcal{A}(I_{a,e^{2\pi t}+a})=U(a,a)\Delta ^{-it}\left( \mathcal{A}%
(W_{a_{+}})^{\prime }\cap \mathcal{A}(W)\right) \Delta ^{it}U^{-1}(a,a)
\end{equation}
where $I_{a,b\text{ }}$ with $b>a\geq 0$ denotes an interval on the right
upper light ray. This net is cyclic and separating with respect to the
vacuum in the reduced Hilbert space 
\begin{eqnarray}
H_{+} &=&\overline{\mathcal{M}_{+}\Omega }=P_{+}H\subset H=\overline{%
\mathcal{A}(W)\Omega } \\
\mathcal{M}_{+} &\equiv &\cup _{t}A(I_{0,e^{2\pi t}}),\,\,E_{+}(\mathcal{A}%
(W))=\mathcal{M}_{+}=P_{+}\mathcal{A}(W)P_{+}  \nonumber
\end{eqnarray}
where the last relation defines a conditional expectation. The application
of $J$ to gives the left lower part of this light ray which is needed for
the full net.
\end{proposition}

\begin{remark}
The most surprising aspect of this proposition is that this light-ray
affiliated chiral conformal theory exhibits the ``blow-up'' property i.e.
can be activated to reconstitute the two-dimensional net by association of
the -light ray translation 
\begin{eqnarray}
\mathcal{A}(W) &=&alg\cup _{a>0}\left\{ \mathcal{M}_{+},U_{-}(a)\right\} \\
\mathcal{A} &=&\mathcal{A}(W)\vee \mathcal{A}(W)^{\prime }  \nonumber
\end{eqnarray}
The Moebius groups SL(2,R)$_{\pm }$ account for 6 parameters in
contradistinction to the 3 parameters of the two-dimensional Poincar\'{e}
group of the massive theory. Most of the former are ``hidden'' and the
original theory perceives these additional symmetries only in its P$_{\pm }$
projections (for the proofs see \cite{Sch-Wie1}\cite{GLRV}).
\end{remark}

The light-ray reduction reduces the derivation of the recursion relation to
a one-dimensional LQP problem and the reader may carry out the missing
algebra without much effort. This reduction also helps significantly in the
demonstration that the $\mathcal{A}(\mathcal{O})$ spaces are non-trivial
i.e. contain more elements than multiples of the identity. It is a
fascinating experience to see that the existence problem for nontrivial
QFT's which in the quantization (Lagrangian, functional integral) approach
always pointed into the direction of getting good short distance properties
and in particular the renormalizability requirement $dim\mathcal{L}%
_{int}\leq dim$spacetime, the modular approach which does not use individual
``field-coordinatizations'' relates the existence of nontrivial field
theories associated with interacting PFG's to the nontriviality of
intersections which represent double cone algebras. The above constructions
only determine operators in the sense of bilinear forms.

At this point it is appropriate to address the question of what we learned
from this approach as compared to the Karowski-Weisz-Smirnov ``axiomatics'' 
\cite{Ka-Wei}\cite{Smir}. Actually a considerable part of that axiomatics
has been reduced to specializations of general field theoretic properties
within the LSZ framework \cite{BFKZ}, apart from the algebraic and analytic
aspects of the fundamental crossing property. Since the LSZ formalism itself
can be derived from the basic causality and spectral properties of say
Wightman QFT, one may even want to have a more direct physical understanding
of the other properties. This is achieved by \ realizing that the a$_{n}$%
-coefficients have the interpretation of the connected part of formfactors
of $A,$ for selfconjugate models 
\begin{eqnarray}
a_{n}(\theta _{1},...,\theta _{n}) &=&\left\langle \Omega \left| A\right|
\theta _{1},...,\theta _{n}\right\rangle ^{in} \\
\theta _{1} &<&\theta _{2}<..<\theta _{n}  \nonumber
\end{eqnarray}
\begin{eqnarray}
&&a_{n}(\theta _{1},..\theta _{\nu },\theta _{\nu +1}-i\pi ,..,\theta
_{n}-i\pi ) \\
&=&^{out}\left\langle \theta _{1},..\theta _{\nu }\left| A\right| \theta
_{\nu +1},..;\theta _{n}\right\rangle _{conn}^{in}  \nonumber
\end{eqnarray}
The relations for different orderings of $\theta ^{\prime }s$ follows from
the algebraic structures of the Z's.

In the diagonal case this connection between Z's and in- and out-
creation/annihilation operators can be seen directly via representing the
Z's in a bosonic/fermionic Fock space of the incoming particles in the form 
\begin{equation}
Z(\theta )=a_{in}(\theta )e^{i\int a_{in}^{\ast }(\theta )a(\theta )d\theta }
\end{equation}

However such representations are not known for the nondiagonal case. But
once one obtained the double cone localized operators the theory itself
(scattering theory as a consequence of the locality+spectral structure)
assures the existence of $Z$ in terms of incoming particle
creation/annihilation operators, albeit not in terms of simple exponential
formulas.

The modular theory for wedges in terms of PFG's really explains the KWS
axiomatics by integrating it back into the fundamental principles of general
QFT. In particular the notoriously difficult crossing symmetry for the first
time finds its deeper explanation in Hawking-Unruh thermal KMS properties
once one realizes that a curved space-time Killing vector (a classical
concept) is not as important quantum localization of operator algebras. With
these remarks we have achieved our goal of deriving and explaining all
axioms of the KWS approach in terms of localization properties of PFG's with
pair interactions.

This raises the question if the PFG's $F(x)$ in their property as wedge
algebra generators, could not exist also for higher dimensions. In that case
their application more than one time to the vacuum would generate state
whose particle content (the real particle structure) is already very
complicated. As often in general QFT, it is easier to see what does not
work, i. e. to prove No-Go theorems. Indeed if the interacting PFG's exist
at all, their causally closed living space $\mathcal{O}$ cannot be (even a
tiny little bit) smaller than a wedge $\mathcal{O}\subset W$. As was already
stated at the beginning, if there would be spacelike directions with an
arbitrarily small conic surrounding which are contained in $W$ but not in $%
\mathcal{O},$ it is fairly easy to generalize the proof of the Jost-Schroer
theorem \cite{ST-Wi} and show that the commutators of such PFG's must be a
c-number which is determined by their two-point function. However the method
used in those No-Go theorems has no extension to the wedge region. If wedge
algebras can indeed be generated by PFG's, one expects again that modular
theory does not only relate them to the S-matrix so that their correlations
can be expressed in terms of products of S-matrix elements and furthermore
that the elusive crossing symmetries for the S-matrix and formfactors find
their explanation in the thermal KMS properties. This surprising relation
between particle physics and the thermal properties of Hawking-Unruh wedge
horizons has attracted the attention of many physicist, the ideas most close
to those of the present work and several older articles \cite{schroer} of
the present author are those in \cite{Nieder}. However it should be clear
that as long as higher dimensional PFG's have yet to be constructed or at
least their existence established, the mediators between off- and on-shell
are still missing and there is no proof beyond the one for factorizing
models presented before.

There is also an interesting extension of the KWS axiomatics in form of a
pair of satellite chiral conformal theories. In contradistinction to the
standard short distance association the light ray association via modular
theory is not just a one way street; the \ blow-up property with the help of
adjoining the opposite light cone translation allows to return, so that
hidden conformal symmetries become relevant for the massive theory or more
precisely for the massive theory projected into the $H_{\pm }$ subspaces.

Note that the present construction principle can be directly used for the
systematic construction of chiral conformal theories. For the construction
of W-like algebras one starts with PFG generators on a half line. Modular
theory assures that in principle every system of S-coefficients fulfilling
the Z-F algebra leads to a bosonic/fermionic conformal theory granted that
the previous relative commutator algebra is non-trivial. This is a
construction scheme which could not have been guessed within the framework
of pointlike fields.

Another apparently simple but untested idea suggested by the present
concepts is the classification of wedge algebras with non-geometric
commutator algebras via statistics Klein factors or constant S-matrices in $%
J.$ Examples are the Ising field theory and the order/disorder fields. For
the more interesting case of plektonic R-matrices which appear in the
exchange algebras \cite{Re-Sch} of charge carrying fields, one knows that
these algebras in contradistinction to bosonic/fermionic (e.g. W-algebras)
are incomplete since the distributional character at coalescent points is
left unspecified. This is not the case if one uses the R-data as an input
into plektonic $Z^{\#}(\theta ).$ The Hilbert space obtained by iterative
application of Z-creation operators is not compatible with a Fock space
structure. Rather the n-particle subspace has the structure of a path space
as known from the representation theory of intertwiner algebras. The
combinatorial complications should be offset by the simplicity of constant
S-matrices. As the operator representation of the massive Ising model shows,
the constant S case should even have a simple coefficient series in the
massive case.

\bigskip

.

\section{\protect\large Concluding Remarks}

Whereas causality and locality principles used to play an important role in
the past (the LSZ framework, the Kramers-Kronig relations in high energy
physics and their experimental check in high energy nucleon scattering),
they have been less prominent in the more global functional integral
formulationof QFT. In S-matrix models as the Veneziano dual model the role
of these principles is even harder to see, but the idea that crossing
symmetry which underlies duality is a deep on-shell manifestation of
causality always carried a lot of plausibility. The difficulty here is that
crossing symmetry was primarily an observation on Feynman diagrams whose
relation to the causality- and particle- structure was never clarified as
that of other symmetries e.g. as it happened with the simpler TCP symmetry.
In fact the dual model which was originally intended to probe the structure
of a nonperturbative S-matrix and to shed light on the elusive crossing
symmetry, was soon treated as a separate issue with the original QFT
motivation being forgotten. After several abrupt changes of interpretation
and finally also of the mathematical formalism (the so called ``string
revolutions'') it finally reached its present form of string theory with
interesting mathematical connections but without convincing conceptual
content. The status of locality within interacting string theory is unknown
(the answer one gets depends on the person asks asks\footnote{%
Part of the problem may originate from the fact that quantum causality and
locality is often confused with support or geometrical properties of
Lagrangians, one of the negative side effects of the naive interpretation of
euclidean field theory.}). If the word string could be interpreted as
indicating a spacetime localization and not just referring to certain
spectral properties, then it would be part of local QFT and all the
structural statements in this article would immediatly be applicable.
However in this case it should be possible to have an intrinsic formulation
(say analogous to the Wightman framework). As it stands now, string theory
is synonymous with a collection of computational steps. Related to this is
the total lack of an answer to the question: what physical principle is it
which asks for a string-like extension in order to be realized? One should
like to have a physically more compelling reason than just saying that after
having been interested for many years in pointlike fields one wants to study
string-like extensions.

The development of physical theories has been (and still is in my opinion)
the unfolding of ever more general realizations of physical principles. For
example the semiinfinite stringlike localization of d=2+1 anyons/plektons or
topological charges (in the sense of algebraic QFT \cite{Haag}) is requires
by the more general realization of causality; if one allows only compact
extensions, one would fall back on bosons/fermions and ordinary charges.
Most structural properties in LQP have been understood as an unfolding of
realizations of physical principles. One hopes that this fruitful viewpoint
of this century may not get completely lost in the ongoing process of
marketing and globalization in the production of publications which is
taking place at the end of it.

\textbf{Acknowledgment: }I am indebted to H. W. Wiesbrock for discussions
and for critical reading of the manuscript.

\BibTeX{}

\end{document}